\documentclass[sigconf,nonacm]{acmart}

\usepackage{subfigure}
\usepackage{tcolorbox}
\usepackage{xspace}
\usepackage{enumitem}
\usepackage{caption}
\usepackage{subcaption}
\usepackage{bbding}
\usepackage{url}
\usepackage{booktabs}
\usepackage{tabularx}
\usepackage{circledsteps}
\usepackage{graphicx}
\usepackage{footmisc}
\usepackage{adjustbox}
\usepackage{pifont}
\usepackage{svg}
\usepackage{xcolor}
\usepackage{caption}
\usepackage{listings}
\usepackage{pifont}

\usepackage{multicol}

\usepackage{makecell}
\usepackage{multirow}
\newcolumntype{C}[1]{>{\centering\arraybackslash}p{#1}}

\usepackage{amsmath,bm,amsthm}
\theoremstyle{definition}

\usepackage{fontawesome}
\usepackage{xcolor}
\usepackage{framed}
\usepackage{textcomp}

\usepackage{microtype}
\newcommand{\code}[1]{\texttt{\footnotesize #1}}

\usepackage[encapsulated]{CJK}

\begin{document}
\begin{CJK*}{UTF8}{gbsn}
% title, conference, authors, keywords, abstract, etc.
\title{When LLMs Meet API Documentation: Can Retrieval Augmentation Aid Code Generation Just as It Helps Developers?}

%% Author info

\author{Jingyi Chen}
\email{jchenix@connect.ust.hk}
\authornote{Co-first authors.}
\affiliation{%
    \institution{The Hong Kong University of Science and Technology}
    \city{Hong Kong}\country{China}
}

\author{Songqiang Chen}
\email{i9s.chen@connect.ust.hk}
\authornotemark[1]
\affiliation{%
    \institution{The Hong Kong University of Science and Technology}
    \city{Hong Kong}\country{China}
}

\author{Jialun Cao}
\email{jcaoap@cse.ust.hk}
\authornote{Corresponding authors.}
\affiliation{%
    \institution{The Hong Kong University of Science and Technology}
    \city{Hong Kong}\country{China}
}

\author{Jiasi Shen}
\email{sjs@cse.ust.hk}
\authornotemark[2]
\affiliation{%
    \institution{The Hong Kong University of Science and Technology}
    \city{Hong Kong}\country{China}
}

\author{Shing-Chi Cheung}
\email{scc@cse.ust.hk}
\affiliation{%
    \institution{The Hong Kong University of Science and Technology}
    \city{Hong Kong}\country{China}
}

%% Abstract

\begin{abstract}
Retrieval-augmented generation (RAG) has increasingly shown its power in extending large language models' (LLMs') capability beyond their pre-trained knowledge. Existing works have shown that RAG can help with software development tasks such as code generation, code update, and test generation. Yet, the effectiveness of adapting LLMs to fast-evolving or \textbf{\textit{less common API libraries using RAG}} remains unknown. To bridge this gap, we take an initial step to study this unexplored yet practical setting -- when developers code with a less common library, they often refer to its API documentation; likewise, \textbf{\textit{when LLMs are allowed to look up API documentation via RAG, to what extent can LLMs be advanced?}} To mimic such a setting, we select four less common open-source Python libraries with a total of 1017 eligible APIs. We study the factors that affect the effectiveness of using the documentation of less common API libraries as additional knowledge for retrieval and generation. Our intensive study yields interesting findings: (1) RAG helps improve LLMs' performance by 83\%$\sim$220\%. (2) Example code contributes the most to advance LLMs, instead of the descriptive texts and parameter lists in the API documentation. (3) LLMs could sometimes tolerate mild noises (typos in description or incorrect parameters) by referencing their pre-trained knowledge or document context. Finally, we suggest that developers pay more attention to the quality and diversity of the code examples in the API documentation. The study sheds light on future low-code software development workflows.
\end{abstract}

%% Keyword


%%
%% This command processes the author and affiliation and title
%% information and builds the first part of the formatted document.
\maketitle

\section{Introduction}
\label{sec:intro}

Retrieval-augmented generation (RAG)~\cite{ragnips20} has recently been introduced as a method to broaden the pre-trained knowledge of Large Language Models (LLMs) by enriching the initial prompts with relevant documents and knowledge via information retrieval~\cite{noise-in-rag}. Since RAG fits perfectly the domains where \textbf{\textit{knowledge is constantly refreshed and can hardly be memorized by LLMs}} in time~\cite{noise-in-rag}, it has increasingly been adopted for software development tasks such as code generation~\cite{lowcode24,su2024arks,gotmare2023efficient,su2024evor,coderagbench} and test generation~\cite{shin2024retrievaltest}.

However, the effectiveness of adapting LLMs to fast-evolving or less common API libraries using RAG remains unknown. Existing studies on RAG-based code generation were made in settings that either mimic updates of common libraries (\textit{e.g.}, SciPy and TensorFlow)~\cite{su2024evor} or explore the upper limits of RAG enhancement in programming problems (\textit{e.g.}, HumanEval~\cite{humaneval} and LeetCode~\cite{livecb})~\cite{coderagbench}. These settings, however, do not study the LLMs' coding capability using less common libraries for software development. 

To bridge the gap, this paper discusses an unexplored yet practical setting: \textbf{\textit{when LLMs generate code using less common libraries, to what extent can RAG contribute?}} {This setting imitates a scenario when human developers use a less common library, they often code by referencing its API documentation~\cite{whyAPIhardtolearn}.
To mimic such a setting,} we select four less common open-source Python libraries with a total of \textbf{1017} eligible APIs. We study the factors that affect the effectiveness of using the documentation of less common API libraries as additional knowledge for \textbf{\textit{retrieval}} and \textbf{\textit{generation}}. Specifically, we study the following five research questions (RQs).

\label{sec:rqs}

\textbf{RQ1: To what extent can the retrieved API documents augment initial prompting and thus enhance the LLMs' generation?} {We consider three scenarios: \ding{202} \textit{Worst case:} {prompting without API documentation}, \textit{i.e.}, imitating developers coding with less common libraries without referencing API documentations; \ding{203} \textit{Best case:} {prompting with only the target API document}, \textit{i.e.}, imitating developers using an API by referring to the correct API documentation; and \ding{204} \textit{Practical case:} {prompting with multiple API documents}, \textit{i.e.}, imitating developers referencing multiple API documents simultaneously when they have not settled down on a specific API.} We also examine the differences in LLM effectiveness across these scenarios and summarize the root causes of LLM failures under each scenario. 

\textbf{RQ2: How do different components in an API document contribute to retrieval and generation?} An API document typically includes function descriptions, parameter lists, and code examples. This RQ explores the contribution of each component to retrieval and augmentation. Also, due to varying document quality, we introduce mild noises in different sections {(\textit{e.g.}, replacing API names in the description and adding a non-existent parameter)} to observe the impact of various noises on retrieval and generation results. {The study aims to provide suggestions on which components of an API document developers should pay more attention to.}

\textbf{RQ3: What is the impact of retrieval methods on retrieval results?} Since RAG acts by searching for relevant queries in external knowledge bases (\textit{i.e.}, API documents in this paper) through information retrieval, the impact of retrieval methods is worth exploring. Intuitively, different retrieval metrics suit different data types (\textit{e.g.}, code, natural language). Yet, it remains unclear which is better for solving coding tasks that involve less common API libraries. 

\textbf{RQ4. How effective is RAG for popular Python libraries?}
In practice, a developer may not be familiar with all APIs in popular libraries. Likewise, LLMs may not be well trained with all APIs of a popular library, particularly the less commonly used ones. Therefore, this RQ examines RAG's effects on retrieving and generating 139 APIs within a popular library, Pandas.

\textbf{RQ5. How does the effectiveness of RAG change when the example code provided in the API document fails to adequately cover the intended usage scenario?} In many cases, developers may find only a basic code example that cannot cover the intended usage scenario of the interested API \cite{whyAPIhardtolearn,usableapis1998,apilearning}. This RQ studies how different examples of an API (\textit{e.g.}, code examples with different parameters or using default parameters) affect the effectiveness of the generation.

Our study yields several interesting findings. 
First, RAG can effectively advance LLMs to correctly use less common APIs by 83\% $\sim$ 220\%. Besides, example code contributes the most to augment LLMs API usage, aligning with their helpfulness to human developers' \cite{whyAPIhardtolearn}. It is also found that LLMs can tolerate mild noises by referencing common APIs and other contexts in documents. In addition, we identify BM25 as an effective retriever for matching the code completion task to the useful API documents. We reveal that RAG also benefits LLMs on the popular Pandas library. Finally, we suggest that developers prepare diverse code examples in API documentation and enhance LLMs' reasoning to avoid insufficient example codes that mislead LLMs. These findings share the experience of enabling LLMs to code with less common APIs, shedding light on the future low-code software development workflow~\cite{lowcode24}. 

\textbf{Our contribution is summarized as follows:}
\begin{itemize}[leftmargin=*]
    \item \textbf{Novelty --} We present the first study on the impact of RAG in the setting where LLMs are leveraged to solve coding tasks involving less common libraries. The setting yields pragmatic observations for AI-assisted software coding.

   \item \textbf{Significance --} We formulate how to use API documentation in RAG for code generation. The intensive experimental results yield interesting findings and actionable advice for both researchers and developers. Our study sheds light on future low-code software development workflows. 
       
    \item \textbf{Usefulness --} We identify the dominating component, \textit{i.e.}, code examples, in the API documentation that contributes the most to the generation. We also suggest prioritizing the importance of different components within the API documentation, allowing developers to focus more on the most useful parts for RAG.
\end{itemize}
\section{Retrieval-Augmented Generation}\label{sec:background}
\label{subsec:ragbg}

Retrieval-Augmented Generation (RAG) \cite{ragnips20} is proposed to enable LLMs to handle tasks by leveraging external knowledge without fine-tuning LLMs. Specifically, RAG maintains a database of relevant domain-specific or latest information unfamiliar to LLMs. Given a query, RAG first runs \textit{retrieval phase} to dynamically identify multiple pieces of information (also known as entities) relevant to the task from the database. Then, the Top-\textit{k} entities are included as the context in the prompt to \textit{augment the generation phase}. 

In the context of documentation-based API usage recommendation, RAG works for a coding requirement query $q$ on a database of API documents $\mathbb{D}=\left\{d_1, d_2, ..., d_n\right\}$, where $d_i$ represents the document of an API unfamiliar to the LLM. It executes two phases:

\textbf{Retrieval Phase.}  
Given the query $q$ as the input to a RAG system, the retriever computes similarity scores between $q$ and each document $d_i\in\mathbb{D}$ to rank all documents. 
The RAG system adopts a parameter $k$ that determines the number of top-ranked documents to use as relevant contextual information. 
The retrieval process can be formally represented as:
\begin{equation*}
\begin{aligned}
\textsc{Retriever}\left(\mathbb{D}, q, k\right) = \textrm{top}\left(\textrm{sorted}\left(\left\{d_1, d_2, ..., d_n\right\}, \textrm{sim}\left(q, \cdot\right)\right), k\right)
\end{aligned}
\end{equation*}

\textbf{Augmented Generation Phase.} A prompt is constructed based on top-$k$ relevant documents as the contextual information and $q$ as the query, with an instruction to clarify the logical relationship between them.
The prompt is input to the generator LLM to get the recommended API usage, which can be formally represented as:
\begin{equation*}
\begin{aligned}
\textrm{Result} \gets \textsc{GeneratorLLM}\left(\left< 
\textsc{Retriever}\left(\mathbb{D}, q, k\right)
, q\right>\right)
\end{aligned}
\end{equation*}

\section{Study Design}\label{sec:studydesign}

\begin{figure*}[h]
\centering
\includegraphics[width=0.9\textwidth]{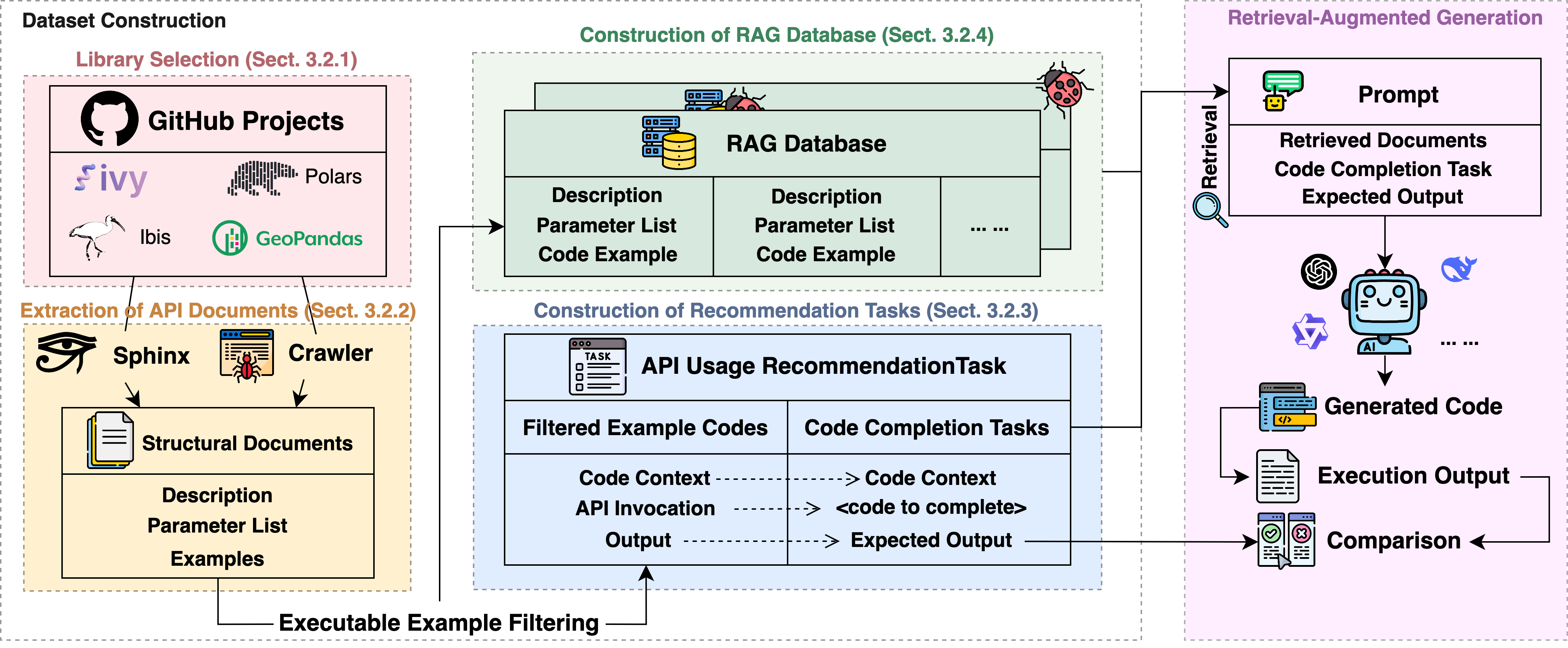}
\caption{Overview of Study Pipeline}
\label{fig:major}
\end{figure*}

\subsection{Overview of Study}\label{subsec:overview}

Figure~\ref{fig:major} illustrates the pipeline of our study. 
We first construct a dataset of API usage recommendation tasks for the study. Specifically, we identify eligible Python libraries as subjects from GitHub to represent the libraries unfamiliar to LLMs (Section~\ref{subsubsec:subjectlibraries}). Then, we extract API documents of the subject libraries into structural data (Section~\ref{subsubsec:docextraction}). Such data are used to construct the RAG database (Section~\ref{subsubsec:constructionrag}) and formulate code completion tasks (Section~\ref{subsubsec:defpbe}). 

Then, we leverage LLMs to complete these tasks using RAG and analyze their performance under different scenarios. Specifically, we construct the prompt for each task by specifying the code context and the expected output, with a succinct task instruction and the retrieved documents attached.
The prompt follows that proposed by existing studies on code generation and RAG \cite{ase24MRAdopt,coderagbench,empiricalofrag}.
The generated codes are evaluated against the consistency of their execution results with the ground truth.
All the prompts and scripts used by our study can be found in our released artifact \cite{artifact}.

% ========================================================

\subsection{Dataset Construction}\label{subsec:datasetconstruction}

\subsubsection{Library Selection} \label{subsubsec:subjectlibraries}

We collect open-source Python libraries as subjects in our study due to the inaccessibility of representative proprietary libraries. Besides, considering the potential volatility of the new libraries and their documents, we select the relatively less common libraries with certain popularity on GitHub as subjects, whose popularity suggests a certain quality of code and documentation.
To this end, we use the 3.8k-starred \textit{best-of-python} list \cite{bestpy} and 19.8k-starred \textit{best-of-ml-python} list \cite{bestpyml} on GitHub as the library source. They list well-recognized Python open-source libraries for various domains and the popular machine learning area, respectively.
Then, we select the subject libraries with the following criteria:

    \textbf{\ding{202} \textit{The libraries should provide structural documents illustrating API description, parameter list, and example codes.}}
    We observe that library API documents typically include a natural language description, a parameter list, and example codes \cite{pydocguideline}. We collect libraries whose documents demarcate the three components with clear sections. Such structural documents facilitate our investigation of the impact of different content in API documents.

    \textbf{\ding{203} \textit{The libraries should include example codes for APIs.}}
    We build target coding tasks based on example codes in API documents. 
    To prepare diverse tasks for study, we only keep the libraries that include more than 100 APIs with at least one coding example.
    
    \textbf{\ding{204} \textit{The library should not rely on a complex runtime environment and its API outputs can be given in text.}}
    We prioritize the libraries that can run without a complicated runtime environment like specific hardware and a server to communicate.
    Besides, in this work, we formulate API usage recommendation as a code completion task and rely on the text representation of the output object to efficiently describe the expected execution result (which will be introduced in Section~\ref{subsubsec:defpbe}). 
    Thus, we only keep the libraries whose API output objects have a clear textual representation. {We discard the libraries whose API outputs are often hard to represent with clear text (\textit{e.g.}, binary object, images, and no output).}

    \textbf{\ding{205} \textit{The library should be used in less than 100k GitHub code files.}}
    The criterion helps select libraries less common in open-source codes; thereby, they tend to be unfamiliar to LLMs.
    Specifically, we use a GitHub regular expression query ``\code{lang:Python /(((import)|(from)).*[\textbackslash s,]<library-name>[\textbackslash s,])/}'' to search the Python code files importing the library. We keep the libraries with fewer than 100k query results. LLMs tend to be less familiar with such libraries. In comparison, popular libraries used in existing studies \cite{ccwanragstudy} can be matched to millions of code files (\textit{e.g.}, \textit{Panads}: 5.3M, \textit{NumPy}: 13.8M, \textit{PyTorch}: 5.5M, \textit{Scikit-learn}: 1.6M).

The first three criteria help select libraries that provide enough eligible data; the last helps select less common libraries. 
Following these criteria, we select four libraries, \textit{i.e.}, \textit{Polars} \cite{polarsrepo}, \textit{Ibis} \cite{ibisrepo}, \textit{GeoPandas} \cite{geopandasrepo}, and \textit{Ivy} \cite{ivyrepo}. 
The first three are data analysis utilities: 
\textit{Polars} and \textit{GeoPandas} are designed for common and geospatial data, respectively, and \textit{Ibis} works as an abstraction layer over various SQL and DataFrame backends. 
\textit{Ivy} unifies APIs of machine learning libraries like PyTorch and TensorFlow \cite{ivy}.

\subsubsection{Extraction of API Documents} \label{subsubsec:docextraction}

\begin{figure}[t]
\centering
\includegraphics[width=0.45\textwidth]{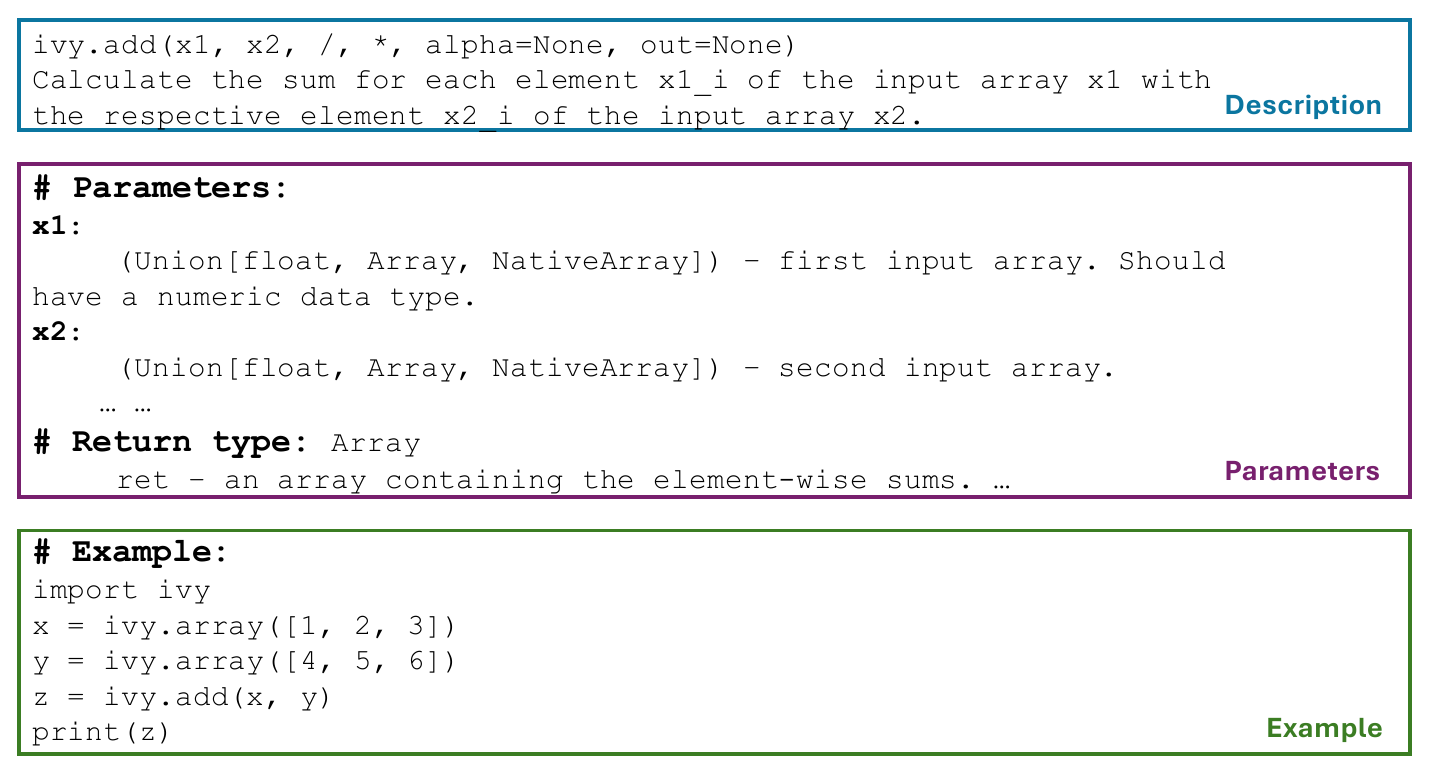}
\setlength{\abovecaptionskip}{5pt}
\caption{Extracted API Document Content of \texttt{ivy.add} API\protect\footnotemark}
\label{fig:opr}
\end{figure}

\footnotetext{Extracted from the Ivy API documentation at \url{https://www.docs.ivy.dev/docs/functional/ivy/elementwise/ivy.functional.ivy.elementwise.add.html\#ivy.add}.}
\begin{figure}[t]
\centering
\includegraphics[width=0.45\textwidth]{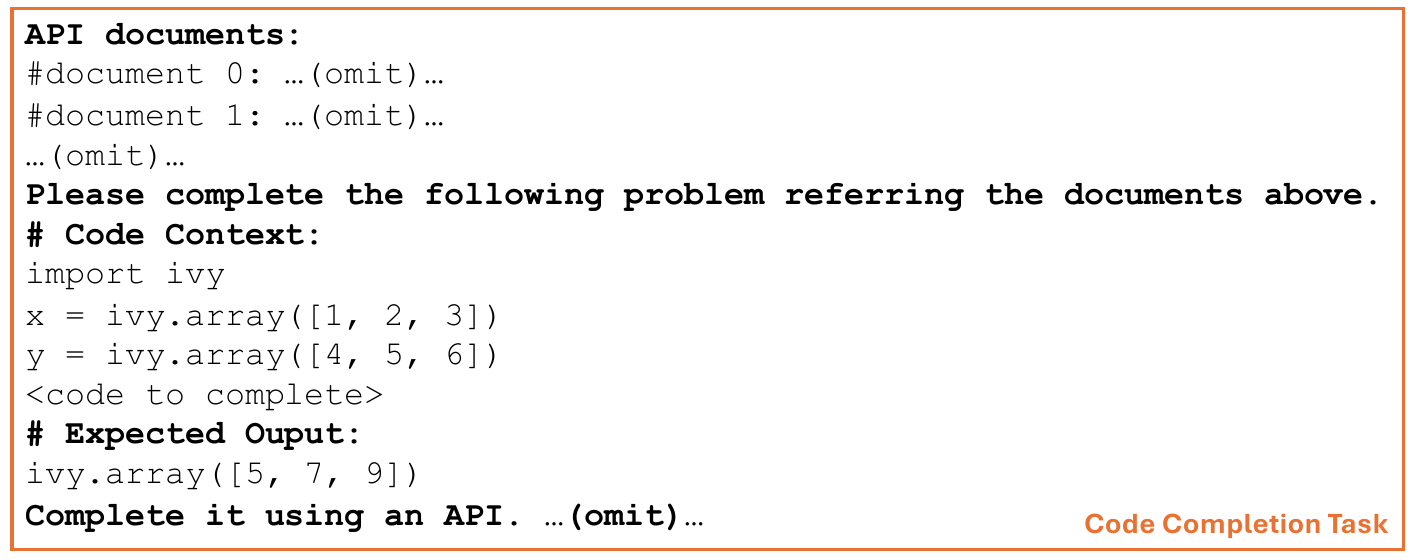}
\setlength{\abovecaptionskip}{5pt}
\caption{A Code Completion Task for \texttt{ivy.add}}
\label{fig:task}
\end{figure}

To build the RAG database and formulate code completion tasks, we parse the API documents of the subject libraries and store them in a structural format. 
Specifically, for each API of a given library, we extract the \textit{description}, \textit{parameter list}, and \textit{example code} from its document. Figure~\ref{fig:opr} shows an extracted API document including the three components, with each annotated in a box. The \textit{description} contains a function signature of the API and several natural language sentences to describe the API function. 
The \textit{parameter list} elaborates all the input parameters and output values of the API, with their function, type, and default value (if any) introduced. 
The \textit{example codes} illustrate typical usage of the API with code snippets, where the code often initializes relevant input values and invokes the API with certain parameter values to realize representative functions. 
We implement a crawler and a parser to collect the API description, parameter list, and example codes from the official API documentation of the subject libraries.

\subsubsection{Construction of Recommendation Tasks}\label{subsubsec: buildtasks} 

\label{subsubsec:defpbe}
In this work, we formulate API usage recommendation as a {code completion task}. 
Specifically, we provide LLMs with the code context (\textit{e.g.}, several code lines to prepare the inputs of invocation) and the expected output, and require LLMs to complete the code snippet with an invocation of the API from a library.
LLMs should suggest an API invocation whose execution result conforms with the expected output to pass the code completion task. 
We consider code completion since it is a common scenario of LLM-assisted development \cite{DeprecatedAPIs}. Besides, code completion tasks can be easily built based on example codes in API documentation and do not require additional information like authentic natural language specifications that may not be easy to collect for less common libraries. 
The expected output is given in the prompt to clarify the intended function following \citet{ds1000} and the idea of Programming-by-Example \cite{pberong}.

Following the idea, we construct {code completion tasks} based on example codes we collected from documents. 
Specifically, given an example code from the API document, we first extract the code line invoking the target API as the goal for LLMs to complete. Then, we collect the codes before the goal as code context. The code lines after the invocation of the target API are removed since the execution result of the target API invocation is generally adequate for assessing the correctness of the suggested API usage on our subjects. 
Besides, to prepare the expected output for {the code completion tasks}, we run the whole example code and record the textual representation of the output object. As Section~\ref{subsubsec:subjectlibraries} mentions, the output of our subject libraries' APIs can be represented in text strings. The strings are included in the prompt to LLMs along with the code context. 

Figure~\ref{fig:task} illustrates a code completion task (and the prompt skeleton) we build based on the example code in Figure~\ref{fig:opr}.
Specifically, we collect the first three lines from the example code as the code context and extract the fourth line as the ground truth.
Executing the whole example code yields \code{z}'s value ``\code{ivy.array([5, 7, 9])}'', which we include in the prompt as the expected output.
The prompt follows the design in existing studies that generate codes based on expected output with LLMs \cite{ase24MRAdopt}. 
The prompts used in our study are available at our artifact \cite{artifact}.

During the construction of tasks, we make necessary edits to make the code examples executable, \textit{e.g.}, inserting \code{import} statements. 
We remove the tasks where the target APIs are not executable after a quick fix or do not produce any output able to be represented in text. 
All remaining completion tasks are checked to be executable and deterministic via multiple executions.
Besides, when the target API appears in multiple statements in one code example, we manually split them into multiple examples. We formulate one {code completion task} per API based on a randomly picked example.

\subsubsection{Construction of RAG Database} \label{subsubsec:constructionrag} 

We build an individual RAG database for each library. Each entity in the database is the document of an API in the library. Each API corresponds to only one document (entity) in the database. 
To study the impact of noises on RAG performance (RQ2), we build a new database with the mutated documents for each library. 
Besides, as introduced in Section~\ref{sec:rqs}, for RQs1-4, we store documents with the example used for the {code completion task} to explore the upper limits of RAG following \citet{coderagbench}; for RQ5, we use another example code that produces an output inconsistent with the code completion task. 
The documents in RAG do not record the output of example codes.

\subsubsection{Statistics} Table~\ref{tab:libstat} summarizes the features of the constructed dataset for our study. It includes four subject libraries with 1017 eligible APIs in total. We extract 2031 example codes for the eligible APIs, with most API documents providing only one example code. As Section~\ref{subsubsec:constructionrag} mentions, we provide one example in each document in the RAG database. The dataset is constructed based on the latest version of the libraries in 2025-02. We release the full dataset in our artifact \cite{artifact}.

\begin{table}[t]
\small
\setlength{\belowcaptionskip}{0pt}
\caption{Statistics on Subject Libraries\label{tab:libstat}}
\resizebox{0.98\linewidth}{!}
{
\setlength{\tabcolsep}{2.4pt}
\begin{tabular}{cccccc} \toprule
\textbf{Library} & \textbf{Stars} & \textbf{\begin{tabular}[c]{@{}c@{}}Matched\\ import\end{tabular}} & \textbf{\begin{tabular}[c]{@{}c@{}}Eligible\\ APIs\end{tabular}} & \textbf{\begin{tabular}[c]{@{}c@{}}Eligible\\ Example Codes\end{tabular}} & \textbf{\begin{tabular}[c]{@{}c@{}}Doc Tokens\\ (per API)\end{tabular}} \\ \midrule
Polars \cite{polarsrepo}    & 32.3k          & 42.8k                                                                 & 330                                                               & 341                                                                & 206.4                                                                 \\ Ibis \cite{ibisrepo}      & 5.6k           & 14.8k                                                                 & 330                                                               & 469                                                                & 140.2                                                               \\ GeoPandas \cite{geopandasrepo} & 4.7k           & 84.5k                                                                 & 119                                                               & 206                                                                & 309.4                                                               \\ Ivy \cite{ivyrepo}       & 14.0k          & 30.5k                                                                 & 238                                                               & 1015                                                               & 279.1                                                               \\ \midrule
\textit{(total)}    & \textit{-}          & \textit{-}   & \textit{1017}  & \textit{2031}   & \textit{-} \\ 
\bottomrule
\end{tabular}
}
\end{table}

% ========================================================

\subsection{Selection of Retrievers and LLMs}

\textbf{Retrievers:} 
We consider three representative retrieval methods, including a sparse retriever, \textit{\textbf{BM25}} \cite{bm25retriever}, and two retrievers based on dense embeddings, text-embedding-3-large (\textit{abbr.} \textit{\textbf{Text3}}) \cite{textembedding} and gte-Qwen2-7B-instruct (\textit{abbr.} \textit{\textbf{GTE}}) \cite{gteembedding}. 
BM25 identifies relevant documents based on TF/IDF. It has been proposed for code generation and retrieval \cite{ccwanragstudy, empiricalofrag, DroidCoder}. 
Besides, we follow \citet{ccwanragstudy} to retrieve documents with Text3, which embeds a text into a 3072-dimensional vector. We do not use MiniLM as they do as MiniLM can only handle short texts. 
Instead, we supplement GTE, a state-of-the-art open-source model on the MTEB embedding leaderboard \cite{mtebbench}. It embeds a text into a 3584-dimensional vector. 

\vspace{0.05cm}

\noindent \textbf{LLMs:} 
We consider four state-of-the-art LLMs from three families on the BigCodeBench leaderboard \cite{BigCodeBench}.
For GPT family, we adopt \textit{\textbf{GPT-4o-mini}} \cite{gpt4omini} due to its popularity and efficiency. We do not consider GPT-4o since we use it to implement some mutation. 
Besides, we consider three open-source LLMs, \textit{i.e.}, Qwen2.5-Coder-32B-Instruct (\textit{abbr.} \textit{\textbf{Qwen32B}}) and Qwen2.5-Coder-7B-Instruct (\textit{abbr.} \textit{\textbf{Qwen7B}}) from Qwen family \cite{qwen25coder} and DeepSeek-Coder-V2-Lite-Instruct (\textit{abbr.} \textbf{\textit{DSCoder}}) from DeepSeek family. They perform well on BigCodeBench and are deployable on our machines.

% ========================================================

\subsection{Mutation Operators for API Documents}\label{subsec:mutoprs}

RAG shows varying effectiveness on external information of different quality \cite{rgb}. 
To understand the importance of each document component and the harm of various noises, we compare RAG performance on mutated documents with certain noise. We design seven mutation operators to simulate noise in practical development.

\subsubsection{Operators on Document Content}

As Section~\ref{subsubsec:docextraction} mentions, API documents typically record \textit{description}, \textit{parameter list}, and \textit{example code} of APIs. 
However, missing specific content in API documents is a critical and common practical issue \cite{DBLP:conf/icse/Aghajani0LMBLS20}. Besides, different content may provide varying helpfulness for LLMs. 
Thus, we design three mutation operators to remove each content from documents. 

\textbf{Delete Description} \textit{(abbr. DelDesc)}: We delete the \textit{description} from documentation while retaining \textit{parameter list} and \textit{example code}.

\textbf{Delete Parameter List} \textit{(abbr. DelParam)}: We delete the \textit{parameter list}, while retaining \textit{description} and \textit{example code}.

\textbf{Delete Example Code} \textit{(abbr. DelExmpl)}: We delete the \textit{example code}, while retaining \textit{description} and \textit{parameter list}.

\subsubsection{Operators on API Names}

API documents may accidentally mention a wrong name of the API \cite{DBLP:journals/tse/LeeWCK21, detectdocerror}.
To understand the impact of this issue on RAG for API usage recommendation, we design two mutation operators to change the API name in documents. 
Note that we do not change the API name in library codes.

\textbf{Add Prefix} \textit{(abbr. AddPrefix)}: 
We prefix original API names with ``\code{func\_}'' in documents. The new (wrong) API name includes the entire original (correct) API name. %The operator retains the 

\textbf{Replace with Synonyms} \textit{(abbr. ReplSyn)}: We replace the original API name with its synonym. 
The synonym is suggested by the widely-used WordNet library \cite{wordnet}.
For the words that WordNet fails to provide a synonym, we employ GPT-4o to suggest a synonym as a legitimate API name. All API name synonyms suggested by GPT-4o were manually verified for reliability.

Note that incorrect API names may exist in both natural language components (\textit{i.e.}, \textit{description} and \textit{parameter list}) and code component (\textit{i.e.}, \textit{example code}) \cite{detectdocerror}. 
To enable a fine-grained analysis, we separately mutate names in natural language and code components.
Take the document in Figure~\ref{fig:opr} as an example, we apply \textit{AddPrefix} on its natural language components to transform the API name ``\code{ivy.add}'' in the function signature into ``\code{ivy.func\_add}''. We also apply \textit{AddPrefix} on its example code to transform the API invocation ``\code{ivy.add(x, y)}'' into ``\code{ivy.func\_add(x, y)}''.

\subsubsection{Operators on API Parameters}  
 
Improper parameter setup is a common issue in API usage \cite{DBLP:conf/sigsoft/HalfondO08}. Inspired by this, we designed two operators to mutate the parameter information in API documents. 
The operators need to generate appropriate parameter descriptions or suggest a new appropriate parameter list. 
Thus, we implement them with GPT-4o instead of a deterministic method to efficiently handle diverse cases. 
We adopt GPT-4o since it shows strong effectiveness in text revision and is widely used for data preparation\cite{caofmbench, domaineval, selfinstruct, LLMannotation}.

\textbf{Add a Parameter} \textit{(abbr. AddParam)}: We prompt GPT-4o to add a new parameter into the function signature and explain it in the parameter list accordingly. For example, GPT-4o inserts a parameter ``\code{beta}'' for the API in Figure~\ref{fig:opr}, creating the signature \code{ivy.add(x1, x2, /, *, beta, alpha=None, out=None)}'' and insert the explanation ``\code{beta: (Union[int, float])-additional...}'' to the parameter list.

\textbf{Rename Parameters} \textit{(abbr. RenmParam)}: We prompt GPT-4o to redesign and rename all the API parameters across the function signature and parameter list. 
For instance, for the API in Figure~\ref{fig:opr}, GPT-4o transforms its signature to \code{ivy.add(array1, array2, /, *, scalar=None, output=None)}'' and updates the parameter list into (\code{array1: ... array2: ... scalar: ... output: ...}'') accordingly.

Even if parameters exist in \textit{description}, \textit{parameter list} and \textit{example code}, in this work, we only rename parameters in natural language \textit{description} and \textit{parameter list} with GPT-4o. Codes may not tolerate the noise of LLM hallucination and need more precise mutation.

\subsection{Experimental Setup}

\subsubsection{LLM Configuration} 
We run LLMs with a temperature of 0.0 to collect their most-confident response. 
We invoke GPT-4o-mini via its API and run the other open-source LLMs on servers equipped with NVIDIA RTX3090 and RTX4090 GPUs with 24GB VRAM.

\subsubsection{Metrics} \label{subsubsec:metric}

We assess the retrieval result based on the widely-used Recall@\textit{k} rate \cite{ccwanragstudy}. The metric reflects the ratio of tasks where the correct document is ranked within Top-\textit{k}. 

For the final recommendation result, we measure the pass rate, \textit{i.e.}, the ratio of tasks where RAG yields a correct API usage \cite{ccwanragstudy}. 
To assess the correctness of the recommended API usage, we extract the code segment enclosed within triple backticks ({\footnotesize$\textquotesingle\textquotesingle\textquotesingle$}) from LLM response and concatenate it to the code context of the {code completion task}. Then, we execute the whole code snippet and collect the execution result. The LLM is considered to pass the task only if the execution result is consistent with the result of ground truth \cite{RAR}.

\label{subsubsec:codevalidation}

\section{Study Results}\label{sec:studyresults}

\subsection{RQ1: Overall Effectiveness of RAG}\label{subsec:rq1}

\begin{table}[t]
\small
\caption{API Usage Recommendation Pass Rates of LLMs w/o and w/\,referring to API Documents\label{tab:rq1-w-wo-doc-top5}}
\hspace*{-2.4pt}\resizebox{1.02\linewidth}{!}{
\setlength{\tabcolsep}{1.2pt}
\begin{tabular}{c@{\hskip 0pt}c|r@{\hskip 0.3pt}lr@{\hskip 0.3pt}lr@{\hskip 0.3pt}lr@{\hskip 0.3pt}l|r@{\hskip 0.3pt}l}
\toprule
\textbf{LLM}                        & \textbf{Setup}      & \multicolumn{2}{c}{\textbf{Polars}} & \multicolumn{2}{c}{\textbf{Ibis}} & \multicolumn{2}{c}{\textbf{GeoPandas}} & \multicolumn{2}{c}{\textbf{Ivy}} & \multicolumn{2}{|c}{\textbf{Pandas}} \\ \midrule

\multirow{3}{*}{\begin{tabular}[c]{@{}c@{}}Qwen\\ 32B\end{tabular}}          & \textit{w/o doc}     & 0.2545 & \textit{}               & 0.2303 & \textit{}                & 0.2437 & \textit{}               & 0.5084 & \textit{}               & 0.6187 & \textit{}               \\
                           & \textit{w/\,top5 doc} & 0.7667 & \textit{{\footnotesize (+201\%)}} & 0.6697 & \textit{{\footnotesize (+191\%)}}  & 0.6639 & \textit{{\footnotesize (+172\%)}} & 0.8151 & \textit{{\footnotesize (+60\%)}}  & 0.8921 & \textit{{\footnotesize (+44\%)}}  \\
                           & \textit{w/\,tgt doc}  & 0.8848 & \textit{{\footnotesize (+248\%)}} & 0.8758 & \textit{{\footnotesize (+280\%)}}  & 0.7899 & \textit{{\footnotesize (+224\%)}} & 0.8992 & \textit{{\footnotesize (+77\%)}}  & 0.9640 & \textit{{\footnotesize (+56\%)}}  \\

\midrule
\multirow{3}{*}{\begin{tabular}[c]{@{}c@{}}GPT-4o\\ -mini\end{tabular}}          & \textit{w/o doc}     & 0.2606 & \textit{}               & 0.2758 & \textit{}                & 0.2521 & \textit{}               & 0.4076 & \textit{}               & 0.6187 & \textit{}               \\
                           & \textit{w/\,top5 doc} & 0.5758 & \textit{{\footnotesize (+121\%)}} & 0.6273 & \textit{{\footnotesize (+127\%)}}  & 0.6273 & \textit{{\footnotesize (+149\%)}} & 0.7647 & \textit{{\footnotesize (+88\%)}}  & 0.8273 & \textit{{\footnotesize (+34\%)}}  \\
                           & \textit{w/\,tgt doc}  & 0.7576 & \textit{{\footnotesize (+191\%)}} & 0.8515 & \textit{{\footnotesize (+209\%)}}  & 0.8571 & \textit{{\footnotesize (+240\%)}} & 0.9160 & \textit{{\footnotesize (+125\%)}} & 0.9281 & \textit{{\footnotesize (+50\%)}}  \\

\midrule
\multirow{3}{*}{\begin{tabular}[c]{@{}c@{}}DS\\ Coder\end{tabular}}          & \textit{w/o doc}     & 0.1061 & \textit{}               & 0.0485 & \textit{}                & 0.1345 & \textit{}               & 0.3487 & \textit{}               & 0.4532 & \textit{}               \\
                           & \textit{w/\,top5 doc} & 0.6485 & \textit{{\footnotesize (+511\%)}} & 0.4485 & \textit{{\footnotesize (+825\%)}}  & 0.6387 & \textit{{\footnotesize (+375\%)}} & 0.7227 & \textit{{\footnotesize (+107\%)}} & 0.7410 & \textit{{\footnotesize (+63\%)}}  \\
                           & \textit{w/\,tgt doc}  & 0.8818 & \textit{{\footnotesize (+731\%)}} & 0.6152 & \textit{{\footnotesize (+1169\%)}} & 0.8655 & \textit{{\footnotesize (+544\%)}} & 0.9118 & \textit{{\footnotesize (+161\%)}} & 0.9568 & \textit{{\footnotesize (+111\%)}} \\

\midrule
\multirow{3}{*}{\begin{tabular}[c]{@{}c@{}}Qwen\\ 7B\end{tabular}}          & \textit{w/o doc}     & 0.1182 & \textit{}               & 0.2182 & \textit{}                & 0.2353 & \textit{}               & 0.3908 & \textit{}               & 0.5612 & \textit{}               \\
                           & \textit{w/\,top5 doc} & 0.3727 & \textit{{\footnotesize (+215\%)}} & 0.5364 & \textit{{\footnotesize (+146\%)}}  & 0.5714 & \textit{{\footnotesize (+143\%)}} & 0.7269 & \textit{{\footnotesize (+86\%)}}  & 0.7410 & \textit{{\footnotesize (+32\%)}}  \\
                           & \textit{w/\,tgt doc}  & 0.4818 & \textit{{\footnotesize (+308\%)}} & 0.6273 & \textit{{\footnotesize (+188\%)}}  & 0.7395 & \textit{{\footnotesize (+214\%)}} & 0.9034 & \textit{{\footnotesize (+131\%)}} & 0.7914 & \textit{{\footnotesize (+41\%)}}  \\

\midrule
\multirow{3}{*}{\textit{(avg)}} & \textit{w/o doc}     & 0.1848 & \textit{}               & 0.1932 & \textit{}                & 0.2164 & \textit{}               & 0.4139 & \textit{}               & 0.5629 & \textit{}               \\
                           & \textit{w/\,top5 doc} & 0.5909 & \textit{{\footnotesize (+220\%)}} & 0.5705 & \textit{{\footnotesize (+195\%)}}  & 0.6253 & \textit{{\footnotesize (+189\%)}} & 0.7574 & \textit{{\footnotesize (+83\%)}}  & 0.8004 & \textit{{\footnotesize (+42\%)}}  \\
                           & \textit{w/\,tgt doc}  & 0.7515 & \textit{{\footnotesize (+307\%)}} & 0.7424 & \textit{{\footnotesize (+284\%)}}  & 0.8130 & \textit{{\footnotesize (+276\%)}} & 0.9076 & \textit{{\footnotesize (+119\%)}} & 0.9101 & \textit{{\footnotesize (+62\%)}} \\

\bottomrule

\multicolumn{12}{l}{\footnotesize \textit{Note: Results on less-used Polars, Ibis, GeoPandas, Ivy are discussed in RQ1; Pandas is discussed in RQ4.}}
\end{tabular}
}
\end{table}

\subsubsection{Setup}\label{subsubsec:rq1setup}

We compare the pass rates of LLMs when referencing no external information (\textit{w/o doc}), the top-5 documents retrieved by BM25 (\textit{w/\,top5 doc}), and the document of the target API (\textit{w/\,tgt doc}). 
As introduced in Section~\ref{sec:intro}, 
setup \textit{w/o doc} simulates the \textit{worst case} where LLMs directly suggest recommendations based on pre-trained knowledge. 
Setup \textit{w/\,top5 doc} simulates the \textit{practical case} of an end-to-end RAG. We consider Top-5 documents followed by {the setup suggested in earlier studies \cite{coderagbench,awsragstudy,ccwanragstudy}}. 
We use BM25 in this RQ since it is robust in domain adaptation \cite{bm25retriever} and found most effective in our study (discussed in RQ3 in Section~\ref{subsec:rq3}).
We also include setup \textit{w/\,tgt doc} to simulate the \textit{best case} where we give LLMs the document of target API to reveal the performance based on an ideal retriever that accurately retrieves correct documents following \citet{coderagbench}.

\subsubsection{Usefulness of Documents in RAG}

As the last three rows in Table~\ref{tab:rq1-w-wo-doc-top5} show, the average pass rate of LLMs improves significantly by 
83\%$\sim$220\% 
after referring to the Top-5 documents retrieved with BM25 (\textit{w/\,top5 doc}) compared to not using RAG (\textit{w/o doc}) on Polars, Ibis, GeoPandas, and Ivy. Specifically, the stronger Qwen32B and GPT-4o-mini suggest correct API usage for 
66\%$\sim$82\% and 58\%$\sim$76\% tasks, respectively. Qwen7B works worst in general, but it still solves more than 53\% of tasks except on \textit{Polars}. The results demonstrate that {\textbf{RAG allows LLMs to learn API usage from the retrieved documents and complete code correctly.}}.

\textbf{Upon Ideal Retrieval:} 
We also analyze the pass rates under setup \textit{w/\,tgt doc}, which reveal RAG performance with an ideal retriever. It is found that Qwen32B, GPT-4o-mini, and DSCoder effectively solve around or over 85\% of tasks most of the time. Such results confirm the helpfulness of RAG for LLMs to learn the correct invocation of unfamiliar library APIs to solve coding tasks.

\subsubsection{Tracing Improvement}

\begin{figure}[t]
\centering
\subfigure[QWen32B]{\includegraphics[width=0.49\linewidth]{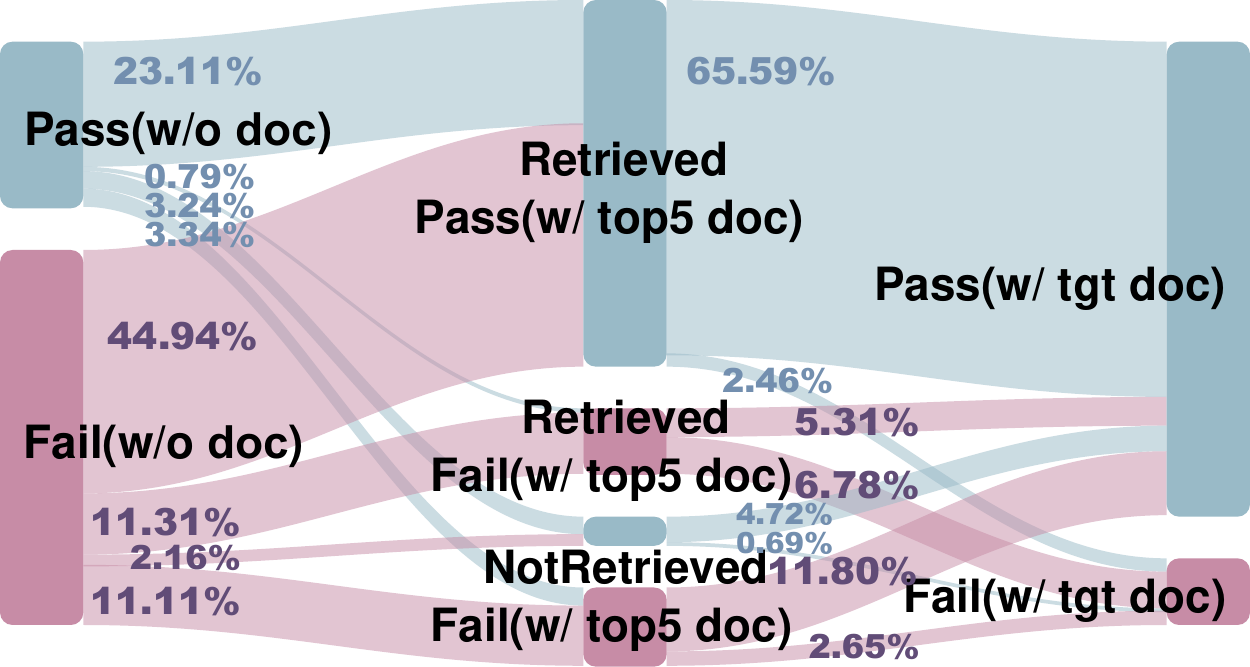}}\hspace{1.4pt}
\subfigure[DSCoder]{\includegraphics[width=0.49\linewidth]{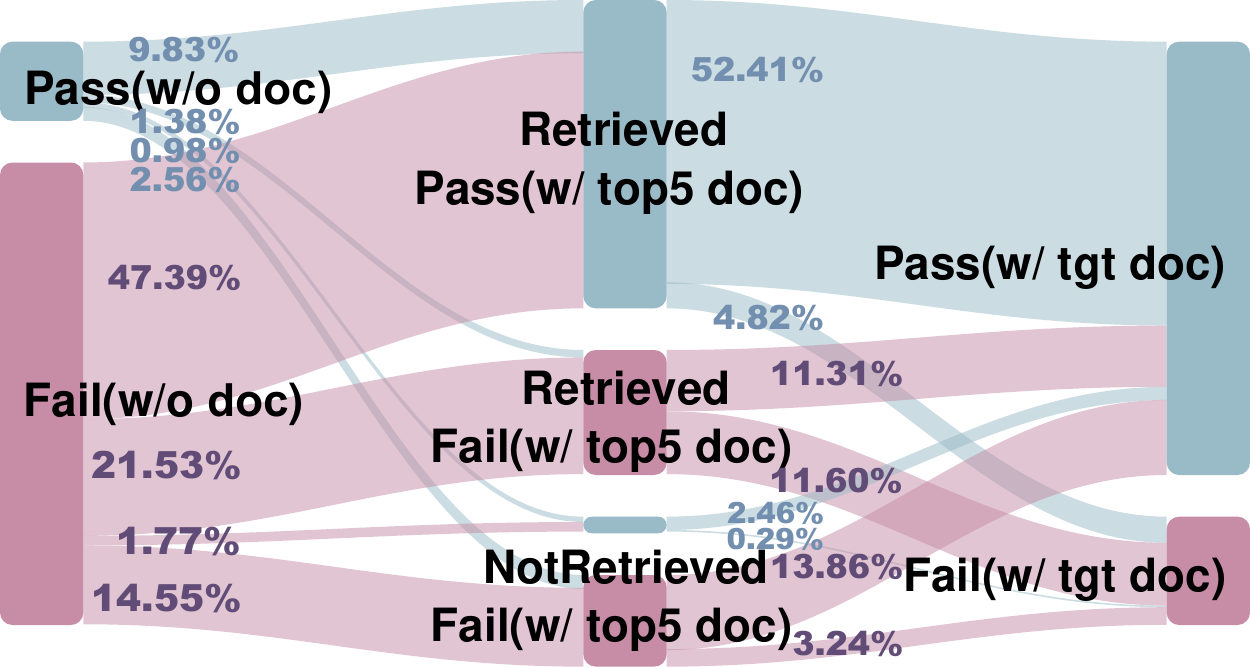}}

\setlength{\abovecaptionskip}{0pt}
\caption{Transition from \textcolor[HTML]{CB748E}{Fail} to \textcolor[HTML]{3D748E}{Pass} on \textit{w/o doc}, \textit{top5 doc}, \textit{corr doc} on Four Less-Common Libraries
\label{fig:rq1-sankey}}
\end{figure}

We trace the passes and failures on different setups to understand how RAG helps LLMs. 
As shown in Figure~\ref{fig:rq1-sankey}, RAG enables LLMs to work out numerous tasks that they fail \textit{w/o doc}. 
Specifically, on 44.94\% and 47.39\% tasks, Qwen32B and DSCoder fail \textit{w/o doc} while succeed \textit{w/\,top5 doc}. 
On setup \textit{w/\,tgt doc}, Qwen32B and DSCoder work out another 11.80\% and 13.86\% tasks, respectively, where they fail due to missing retrieval of correct documents \textit{w/\,top5 doc}.
\textbf{The results again confirm the significant helpfulness of RAG and suggest enhancing retrievers as a direction to enlarge benefit.} 
Meanwhile, RAG hinders successful recommendation on only a few tasks, \textit{e.g.}, 4.13\% (0.79\%+3.34\%) and 3.94\% (1.38\%+2.56\%) where Qwen32B and DSCoder can solve \textit{w/o doc}, respectively. The results on GPT-4o-mini and Qwen7B show consistent conclusions and can be found in our artifact \cite{artifact}.

\subsubsection{Failure Cases}\label{subsubsec:failurecases}

LLMs still fail to solve a few tasks with RAG. For example, Figure~\ref{fig:rq1-sankey} shows that failure of RAG with Qwen32B and DSCoder on 11.80\% and 13.86\% tasks is due to \textit{\textbf{missing retrieval of correct documents}}. Besides, even if the correct API documents have been retrieved within Top-5, LLMs still fail on some tasks. We manually summarized three failure patterns based on these cases, expecting that they may inspire enhancement of RAG on our task.

\ding{202} \textit{LLMs are confused by the retrieved documents and select a wrong API.} The issue causes most failures when the correct document is retrieved within Top-5. For example, for a task of Polars \code{starts\_with} API, BM25 additionally retrieves \code{ends\_with} and another three APIs and Qwen32B wrongly selects \code{ends\_with}. But it solves the task on setup \textit{w/\,tgt doc}. 
The issue may be tackled by enhancing retrieval accuracy to reduce noisy context and improving LLMs in API selection.

\ding{203} \textit{LLMs hallucinate a wrong API usage.} Even on setup \textit{w/\,tgt doc}, weaker LLMs may hallucinate an incorrect API usage. For example, for a Polars task expecting \code{s.rolling\_quantile(quantile=0.33,window \_size=3)}, DSCoder wrongly returns \code{s.rolling\_quantile(quantile=0.33, window\_size=3\underline{, min\_periods=1})}; while Qwen32B succeeds. Enhancing LLMs in document understanding may help solve the issue.

\ding{204} \textit{LLMs struggle to use APIs with complicated parameters.}  Even on setup \textit{w/\,tgt doc}, LLMs may not set parameters correctly when the API has a few complex parameters. For example, \code{polars.Expr.ewm\_std} has eight parameters. Qwen32B wrongly uses \code{spam} but not the desired \code{com} parameter even if \code{com} is already illustrated in the example. In fact, all studied LLMs cannot use this API correctly. 
A method to help LLMs distinguish complex parameters may solve the issue.

\subsubsection{Implications} We summarize insights based on the observed overall effectiveness of RAG on API usage recommendation.

\ding{202} RAG enables LLMs to {solve code completion tasks by correctly invoking unfamiliar library APIs based on API documents}.

\ding{203} LLMs often use APIs correctly when target API documents are retrieved. 
A few wrong API uses of RAG are due to failing to retrieve documents of correct APIs or being misled by high-ranking documents of irrelevant APIs. Improving retrieval accuracy to rank the target API document higher should effectively enhance RAG.

\ding{204} LLMs may still be confused by candidate APIs and complicated parameters. They should be enhanced towards better discrimination on APIs and understanding of complex API parameter setup.

\subsection{RQ2: Effectiveness of RAG on Documents of Different Quality}\label{subsec:rq2}

\begin{figure*}[t]
\centering
\subfigure[Polars]{\includegraphics[width=0.249\linewidth]{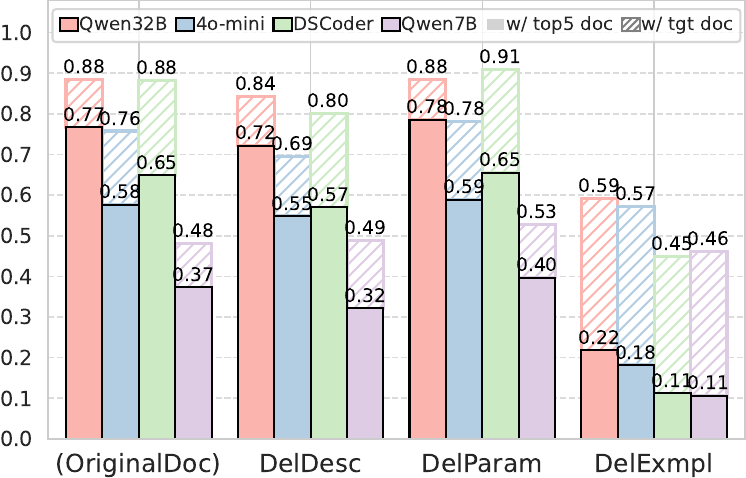}}\hspace{-2pt}
\subfigure[Ibis]{\includegraphics[width=0.249\linewidth]{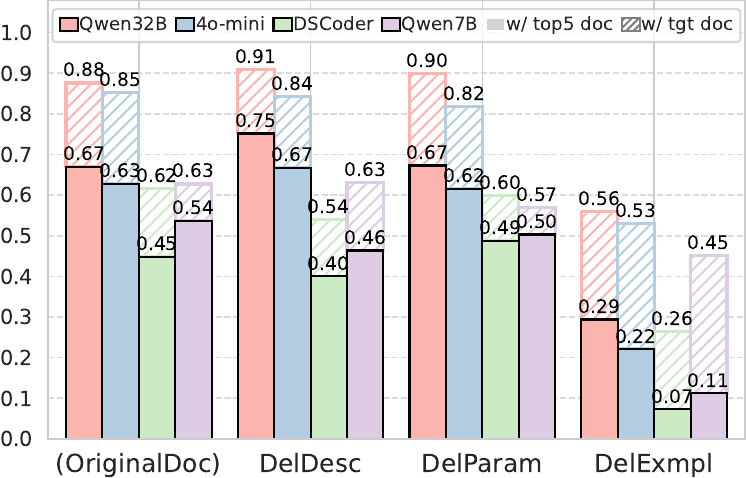}}\hspace{-2pt}
\subfigure[GeoPandas]{\includegraphics[width=0.249\linewidth]{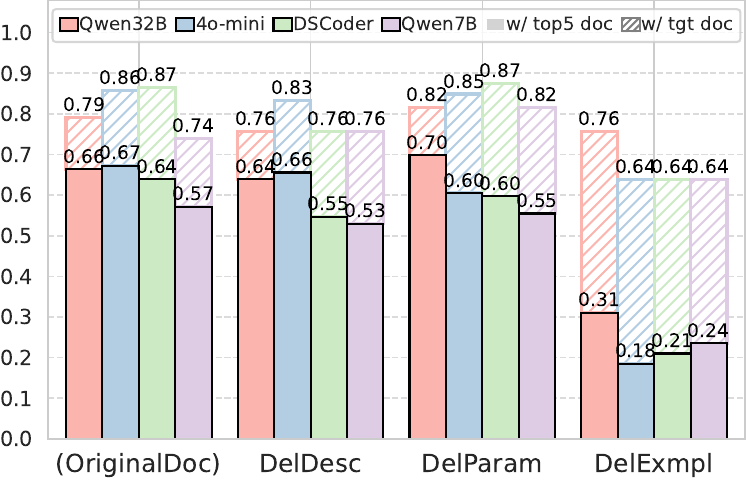}}\hspace{-2pt}
\subfigure[Ivy]{\includegraphics[width=0.249\linewidth]{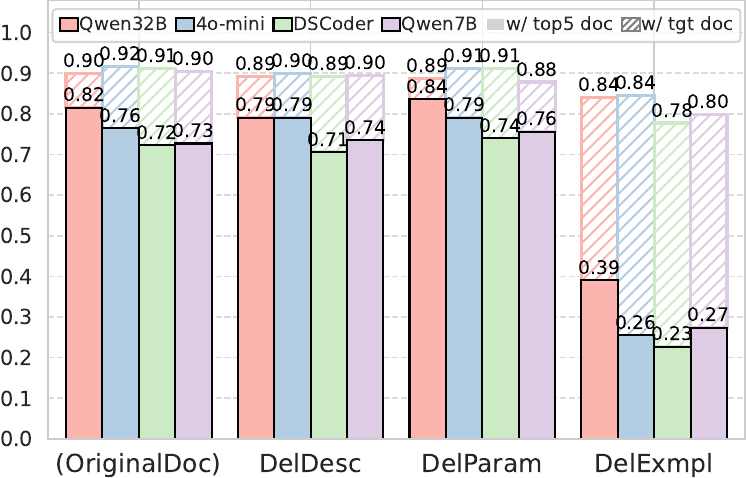}}
\setlength{\abovecaptionskip}{3pt}
\caption{API Usage Recommendation Pass Rates of RAG on Documents Lacking Different Contents (Solid and shaded bars illustrate results under setups w/ top5 doc and w/tgt doc, respectively, as introduced in Section~\ref{subsubsec:rq1setup})\label{fig:rq2contentopr}.}
\end{figure*}

\begin{figure*}[t]
\centering
\subfigure[Polars]{\includegraphics[width=0.249\linewidth]{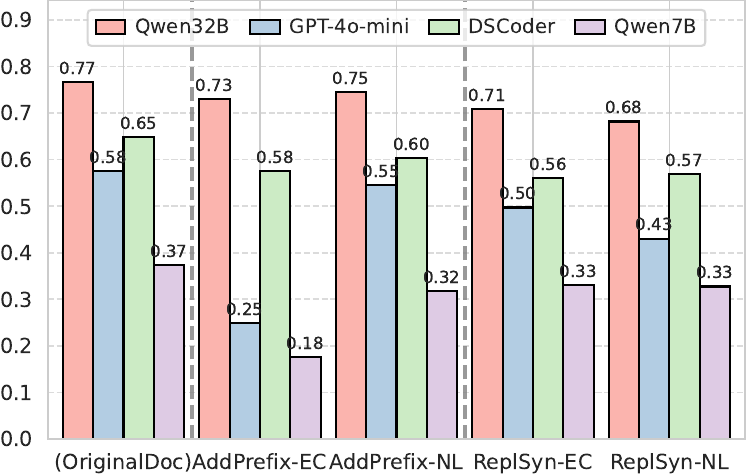}}\hspace{-2pt}
\subfigure[Ibis]{\includegraphics[width=0.249\linewidth]{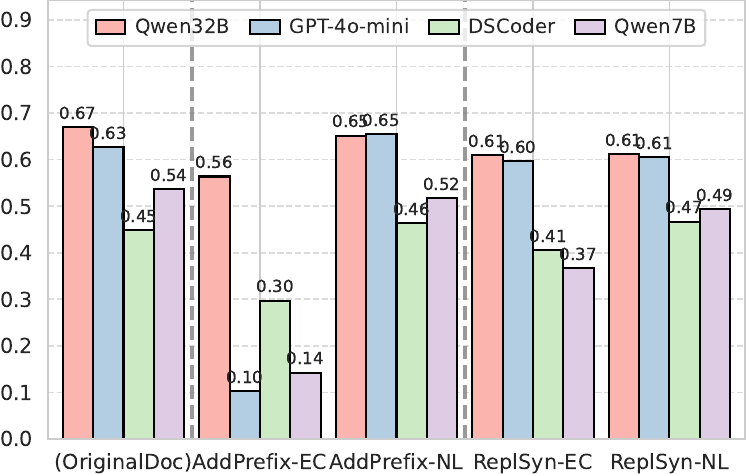}}\hspace{-2pt}
\subfigure[GeoPandas]{\includegraphics[width=0.249\linewidth]{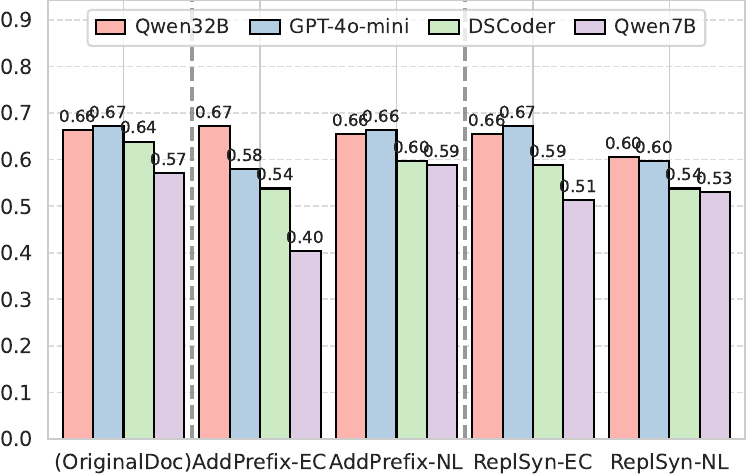}}\hspace{-2pt}
\subfigure[Ivy]{\includegraphics[width=0.249\linewidth]{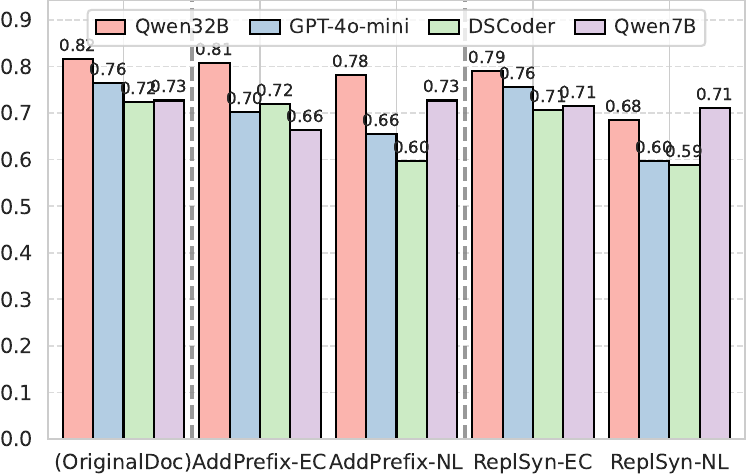}}

\setlength{\abovecaptionskip}{3pt}
\caption{API Usage Recommendation Pass Rates of RAG on Documents with Different Noise on API Names (Postfixes ``-EC'' and ``-NL'' refer to the mutation on example codes and natural language description and parameter list, respectively.)\label{fig:rq2apinameopr}}
\end{figure*}

\begin{figure}[t]
\centering
\subfigure[Polars]{\includegraphics[width=0.49\linewidth]{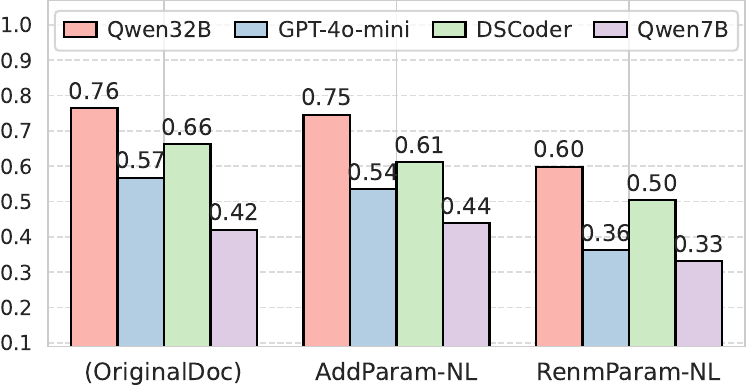}} {\hspace{0pt}}
\subfigure[Ibis]{\includegraphics[width=0.49\linewidth]{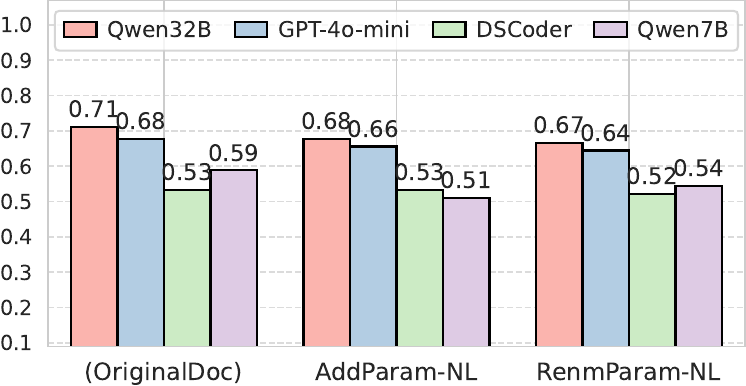}}

\subfigure[GeoPandas]{\includegraphics[width=0.49\linewidth]{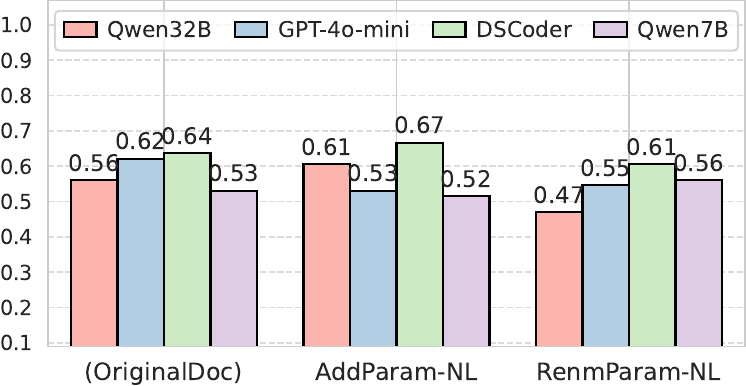}} {\hspace{0pt}}
\subfigure[Ivy]{\includegraphics[width=0.49\linewidth]{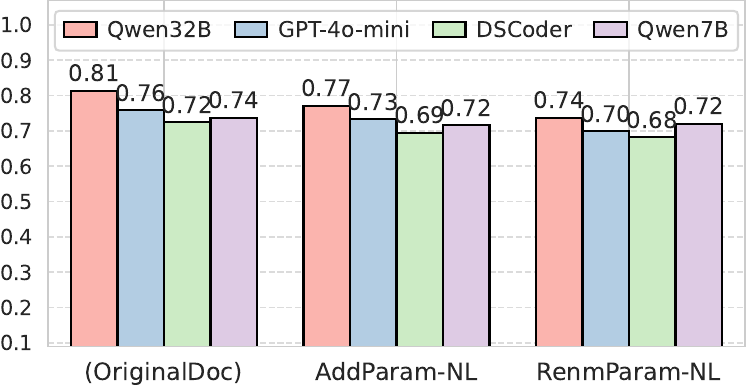}}
\setlength{\abovecaptionskip}{0pt}
\caption{RAG Performance on Documents with Different Noise on Parameters (on the APIs with at least one parameter)\label{fig:rq2paramopr}}
\end{figure}

\subsubsection{Setup} In this RQ, we compare API usage recommendation pass rates of RAG on documents with different noise. We still report the result with the BM25 retriever since it is found to be most effective in general (based on RQ3 results in Section~\ref{subsec:rq3}). 
\subsubsection{Impact of Lacking Content} We first study the performance on the BM25-retrieved Top-5 documents with certain content deleted (solid bars in Figure~\ref{fig:rq2contentopr}). Obviously, the pass rates of LLMs decrease most after we remove example codes (\textit{DelExmpl}) from documents. For example, the pass rates of Qwen32B decrease from {0.66$\sim$0.82} to {0.22$\sim$0.39} on the four libraries. The result shows the essential impact of example codes on the success of the whole RAG system. \textit{Interestingly, the finding aligns with human developers' heavy reliance on example codes to learn API usage \cite{whyAPIhardtolearn}.}

To understand the pure impact on LLM code generation regardless of retrieval phase, we further check LLM performance on target API documents (shaded bars in Figure~\ref{fig:rq2contentopr}). \textit{DelExmpl} also leads to the largest performance drop in this situation, \textit{e.g.}, the pass rates of Qwen32B decrease 
from {0.79$\sim$0.90} to {0.56$\sim$0.84}.
The results confirm the importance of examples for LLMs to learn API usage. 

Meanwhile, lacking description (\textit{DelDesc}) causes only slight performance change. Interestingly, removing the parameter list (\textit{DelParam}) even slightly improves generation (as well as retrieval, which will be discussed in RQ3). 
The reason may be that LLMs can already learn API usage based on the example codes and descriptions, and removing the parameter list alleviates distraction.

\textbf{Project-wise:}
We notice that \textit{DelExmpl} has a slighter hindrance of RAG on Ivy than on the other three libraries. 
This is likely because Ivy APIs mirror the functions in popular libraries like PyTorch and TensorFlow since Ivy aims to unify the deep learning interfaces \cite{ivy}.
As a result, once the correct Ivy API is identified, LLMs may infer its usage by referencing similar APIs in familiar libraries like PyTorch and TensorFlow. This suggests {LLMs tend to be more robust when using unfamiliar APIs that resemble popular ones}.

\subsubsection{Noises at API Names and Parameters}
In this section, we study RAG performance on documents with noises at API names and parameters. 
We mainly discuss the performance on the BM25-retrieved top-5 documents (\textit{w/\,top5 doc}) in this RQ since the noisy API names and parameter information disturb BM25 slightly (discussed in RQ3) and the results on target API documents show similar trends. 

\textbf{Noise at API Names:} 
{We first compare the impacts of adding a prefix to the API names in the example code (\textit{AddPrefix-EC}) and in the natural language description and parameter list (\textit{AddPrefix-NL}).}
It is found that noisy API names in example codes (-EC) lead to a larger decrease in pass rates. The results suggest LLMs prefer to follow EC when there is a minor inconsistency between API names in NL and EC.
Meanwhile, the impacts differ slightly between \textit{ReplSync-EC} and \textit{ReplSync-NL}, with noise in NL components being slightly more destructive. This indicates that LLMs do not manifest a clear preference for API names in NL and EC when the inconsistent names are quite different in syntax. LLMs may wisely select the correct name based on other descriptive contexts in documents.

\textbf{Noise at Parameter Info:} We observe that the impact of two mutations on parameters in NL components is minor in general. This suggests the robustness of LLMs in tolerating such noise. Specifically, even if we introduce a new parameter, LLMs may wisely learn from the example code that the new parameter is optional to realize the goal. Therefore, \textit{AddParam-NL} may not seriously mislead LLMs. Meanwhile, Python does not necessarily require explicitly mentioning parameter names in invocation; thus the impact of parameter renaming is not dramatic either. However, we observe that LLMs sometimes explicitly mention the names of adopted parameters, leading to a more obvious performance drop on \textit{RenmParam-NL}.

\subsubsection{Implications}

Based on findings on various noisy documents, we summarize advice on API document preparation for RAG.

\ding{202} Example codes are vital to the success of RAG in API usage recommendation, aligning with their helpfulness to human developers \cite{whyAPIhardtolearn,usableapis1998}.
RAG users should provide examples in API documentation to teach LLMs API usage. (RQ5 will further discuss the preparation of example codes.)

\ding{203} LLMs prefer following examples when natural language components conflict with examples slightly. RAG users should carefully examine the quality of example code. Besides, explicitly mentioning optional information in examples, \textit{e.g.}, correct parameter names and values, may guide LLMs to learn correct API usage.

\ding{204} LLMs can tolerate noises in documents by referencing context or APIs familiar to them. Prompting LLMs to verify the answer with various information may enhance their robustness against noise in API documents.

\subsection{RQ3: Performance of Different Retrievers}\label{subsec:rq3}

\begin{figure}[t]
\centering
\subfigure[Polars\label{subfig:rq3retrieveronnoise-polars}]{\includegraphics[width=0.49\linewidth]{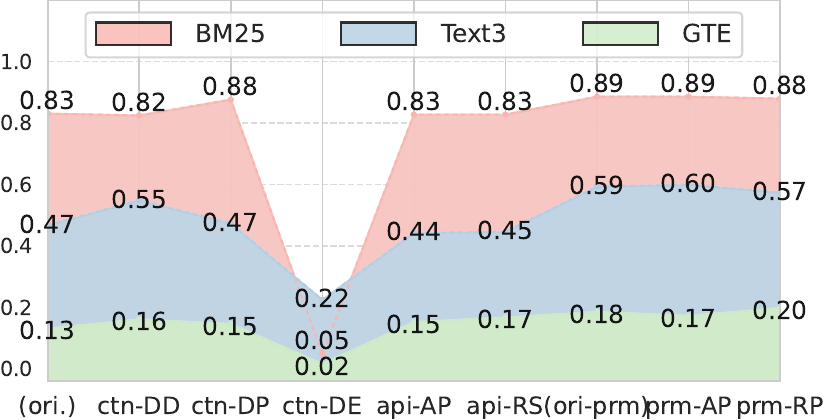}} {\hspace{0pt}}
\subfigure[Ibis]{\includegraphics[width=0.49\linewidth]{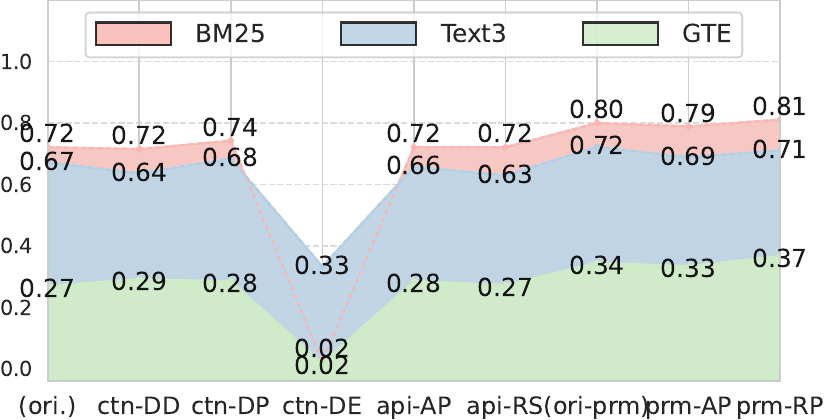}}

\subfigure[GeoPandas]{\includegraphics[width=0.49\linewidth]{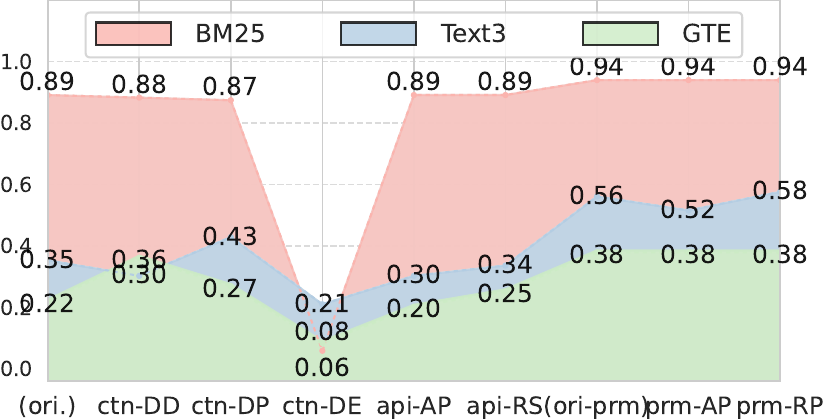}} {\hspace{0pt}}
\subfigure[Ivy]{\includegraphics[width=0.49\linewidth]{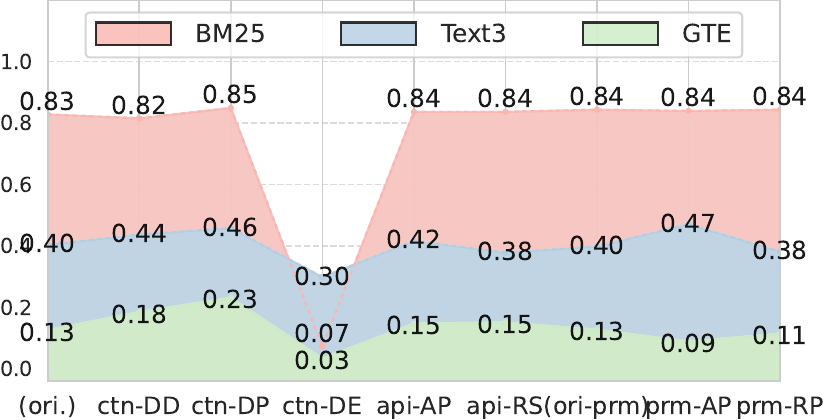}}
\setlength{\abovecaptionskip}{3pt}
\caption{Recall@5 of Retrievers against Different Noise\label{fig:rq3retrieveronnoise}
}
\end{figure}

\subsubsection{Setup} 

We compare retrievers using recall@\textit{k} rate as mentioned in Section~\ref{subsubsec:metric}. We set \textit{k} to be 5 as in earlier RQs. Besides, we also study the Mean Reciprocal Rank (MRR), which provides a single-score summary of retrieval performance across different \textit{k}s. The MRR results show consistent findings and can be found at \cite{artifact}. 
\subsubsection{Comparison of Retrievers} As shown in Figure~\ref{fig:rq3retrieveronnoise}, BM25 shows a generally much higher Recall@5 rate than Text3 and GTE except against deleting examples (\textit{ctn-DE}). Specifically, except on \textit{ctn-DE}, BM25 shows {0.72$\sim$0.94} recall@5 rates;
while Text3 scores only 0.21$\sim$0.72 and GTE performs worse. The result aligns with findings in \cite{empiricalofrag,YeasandNays} that BM25 is effective on code. 
{The reason may be that for the requirement described with code context and expected output, matching codes literally with BM25 is more helpful than considering the semantics Text3 and GTE compress in a vector.}

\subsubsection{Noise-wise Comparison}

As shown in Figure~\ref{fig:rq3retrieveronnoise}, among different noise, deletion of example code content (\textit{ctn-DE}) decreases retrieval accuracy a lot. 
In this case, three retrievers can retrieve correct documents for no more than 33\% tasks within Top-5.
The results reveal the importance of example codes for accurate retrieval. 
Besides, deleting the parameter list (\textit{ctn-DP}) slightly increases the accuracy of all retrievers. The reason may be that the parameter list offers little information about the general features of APIs and may introduce noise by contrast. 

\subsubsection{Implications}

In general, BM25 works most effectively in our setup, {demonstrating that ``{Code Context} + BM25-based RAG'' as a potentially promising solution to API usage recommendation when there are reliable example codes in documents}. This is reasonable since codes are often concise and structured, and thus easy to match to example codes and API information in documents literally.

\subsection{RQ4: Comparison with Popular Library}\label{subsec:rq4}

\begin{table}[t]
\small
\caption{Relative Performance Change against Different Noise on Less-Used Libraries and Popular Pandas Library\label{tab:rq4-vsPandas-noise}}
\hspace*{-2.2pt}\resizebox{1.015\linewidth}{!}{
\setlength{\tabcolsep}{1.5pt}
\begin{tabular}{cc|ccccccccc|c}
\toprule

\textbf{LLM}	 & \textbf{Library}	 & \textbf{\begin{tabular}[c]{@{}c@{}}ctn-\\ DE\end{tabular}}	 & \textbf{\begin{tabular}[c]{@{}c@{}}ctn-\\ DP\end{tabular}}	 & \textbf{\begin{tabular}[c]{@{}c@{}}ctn-\\ DD\end{tabular}}	 & \textbf{\begin{tabular}[c]{@{}c@{}}api-\\ AP (nl)\end{tabular}}	 & \textbf{\begin{tabular}[c]{@{}c@{}}api-\\ AP(ec)\end{tabular}}	 & \textbf{\begin{tabular}[c]{@{}c@{}}api-\\ RS(nl)\end{tabular}}	 & \textbf{\begin{tabular}[c]{@{}c@{}}api-\\ RS(ec)\end{tabular}}	 & \textbf{\begin{tabular}[c]{@{}c@{}}prm-\\ AP\end{tabular}}	 & \textbf{\begin{tabular}[c]{@{}c@{}}prm-\\ RP\end{tabular}} & \textbf{(avg)} \\ \midrule

\multirow{2}{*}{\begin{tabular}[c]{@{}c@{}}Qwen\\ 32B\end{tabular}}
  	 	 	 	       	       & \textit{avg.} 4 libs	 & -58\% & 3\%  & 0\%   & -3\% & -5\%  & -11\% & -5\%  & -2\%  & -13\% & -11\% \\ 
	 & Pandas            	 & -36\% & -2\% & -4\%  & 2\%  & 0\%   & 1\%   & -2\%  & -2\%  & -6\%  & -5\% \\ 
\midrule

\multirow{2}{*}{\begin{tabular}[c]{@{}c@{}}GPT-4o\\ -mini\end{tabular}}	 
  	 	 	 	   	       & \textit{avg.} 4 libs	 & -68\% & 0\%  & 3\%   & -3\% & -37\% & -14\% & -3\%  & -6\%  & -14\% & -16\% \\ 
	 & Pandas            	 & -44\% & 3\%  & 5\%   & -3\% & -1\%  & -8\%  & -6\%  & -3\%  & -4\%  & -7\%  \\ 
\midrule

\multirow{2}{*}{\begin{tabular}[c]{@{}c@{}}DS\\ Coder\end{tabular}}
	 	 	 	 	   	       & \textit{avg.} 4 libs	 & -75\% & 1\%  & -10\% & -8\% & -13\% & -12\% & -8\%  & -2\%  & -10\% & -15\% \\ 
	 & Pandas            	 & -51\% & 1\%  & 7\%   & 1\%  & -7\%  & 1\%   & -7\%  & 5\%   & 0\%   & -6\%  \\ 
\midrule

\multirow{2}{*}{\begin{tabular}[c]{@{}c@{}}Qwen\\ 7B\end{tabular}}
  	 	 	 	   	       & \textit{avg.} 4 libs	 & -67\% & 0\%  & -7\%  & -3\% & -37\% & -7\%  & -13\% & -4\%  & -5\%  & -16\% \\ 
	 & Pandas            	 & -43\% & -4\% & 0\%   & -1\% & -20\% & -2\%  & -5\%  & -11\% & -6\%  & -10\%  \\ 
\bottomrule

 \multicolumn{12}{l}{\footnotesize \textit{Relative Change Rate = $(\textrm{Perf}_{\textrm{NoisyDoc}} - \textrm{Perf}_{\textrm{OriDoc}}) / \textrm{Perf}_{\textrm{OriDoc}}$. Detailed values available at artifact \cite{artifact}.}}\\
 \multicolumn{12}{l}{\footnotesize \textit{``avg. 4 libs'' refers to average results on Polars, Ibis, GeoPandas, and Ivy. Full results available at \cite{artifact}.}} \\
 \multicolumn{12}{l}{\footnotesize \begin{tabular}[l]{@{}l@{}} Prefixes \textit{ctn-}, \textit{api-}, \textit{prm-} refer to operators on contents, API names, and parameters, respectively; \\the last two letters are the initials of operator names.\end{tabular}}
\end{tabular}
}
\end{table}

\subsubsection{Setup} We compare the effectiveness of RAG on the four less common libraries and the popular Pandas library. Pandas matches 5.4M GitHub codes with our query mentioned in Section~\ref{subsubsec:subjectlibraries}. 
It is used as a subject in an existing study on RAG \cite{ccwanragstudy} and for code generation \cite{ds1000}. We extract 139 eligible Pandas APIs for study. 

\subsubsection{Consistent and Generalizable Findings}
We observe some consistent findings on the less common libraries and Pandas. 
Specifically, Table~\ref{tab:rq1-w-wo-doc-top5} shows that RAG also helps improve pass rates on Pandas. For example, Qwen32B works out another 44\% and 56\% tasks with the help of \textit{top5 doc} and \textit{tgt doc}, respectively. 
Besides, Table~\ref{tab:rq4-vsPandas-noise} shows that the removal of example codes (\textit{ctn-DE}) also causes large performance drops of RAG on Pandas, demonstrating the importance of example codes to the success of RAG on Pandas.

The similar findings show the generalizability of our conclusions and implications. RAG together with well-prepared examples also effectively helps LLMs use APIs from well-learned libraries.

\subsubsection{Unique Observations} \label{subsubsec:rq4diff}
Besides consistent findings, we observe differences between results on Pandas and four less common libraries. Specifically, Table~\ref{tab:rq1-w-wo-doc-top5} shows that LLMs without referencing documents (\textit{w/o doc}) can already solve 56.29\% tasks on Pandas on average, which is much higher than 18.48\%$\sim$41.39\% 
on four less common libraries. 
Besides, Table~\ref{tab:rq4-vsPandas-noise} shows that the pass rates of RAG decrease by 11\%$\sim$16\% on four less common libraries while only 5\%$\sim$10\% on Pandas. In particular, RAG is much more robust to noise of wrong API names, notably, \textit{api-AP(ec)} and \textit{api-RS(nl)}. 

The results show that studying RAG on only popular libraries may hide some problems of RAG on less common libraries. They echo our selection of less common libraries as subjects.

\subsection{RQ5: Effectiveness on Documents with Only Example Codes Mismatching Target Tasks}\label{subsec:rq5}

\begin{figure}[t]
\centering
\includegraphics[width=0.72\linewidth]{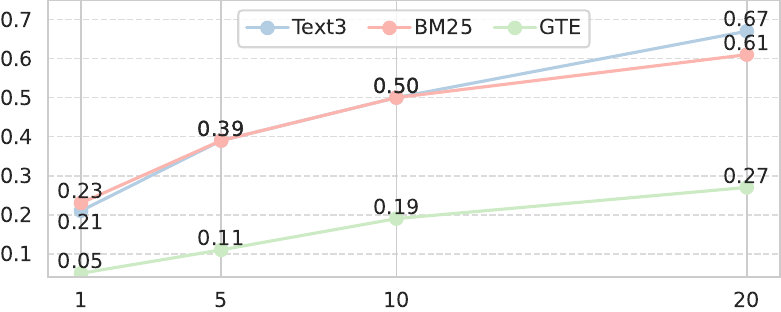}
\setlength{\abovecaptionskip}{4pt}
\caption{Recall@\textit{k} on Tasks Different From Example Code\label{fig:rq5retriever}}
\end{figure}

\begin{table}[t]
\small
\caption{Pass Rates on Tasks Different From Example Code\label{tab:rq5-top1ondifftask}}
\resizebox{0.96\linewidth}{!}
{
\setlength{\tabcolsep}{4pt}
\begin{tabular}{ccccc}
\toprule
           & \textbf{\textit{w/o doc}} & \begin{tabular}[c]{@{}c@{}}\textbf{\textit{w/\,tgt doc}}\\ {\textit{(Original Doc)}}\end{tabular} & \begin{tabular}[c]{@{}c@{}}\textbf{\textit{w/\,tgt doc}}\\ {\textit{(DelExmpl)}}\end{tabular} & \begin{tabular}[c]{@{}c@{}}\textbf{\textit{w/\,tgt doc}}\\ {\textit{(OnlyExmpl)}}\end{tabular} \\ \midrule
Qwen32B    & 0.5564  & 0.8346      & 0.8421      & 0.7744       \\
GPT-4o-mini& 0.4361  & 0.7895      & 0.8271      & 0.7444       \\
DSCoder    & 0.4060  & 0.6917      & 0.7970      & 0.6165       \\
Qwen7B     & 0.4436  & 0.6617      & 0.7820      & 0.6241       \\ \midrule
R1-Qwen32B & 0.5789  & 0.8496      & 0.8346      & 0.7444       \\ \bottomrule
\end{tabular}
}
\end{table}

\subsubsection{Setup} 

This RQ explores the effectiveness of RAG on API documents whose example code differs from the {code completion task}, aiming to shed light on the preparation of example codes in API documents for RAG. 
However, existing documents seldom provide multiple examples for each API \cite{whyAPIhardtolearn}. Among our subjects, only Ivy offers more than one example for over 100 APIs. Thus, we study this RQ on the 133 Ivy APIs with multiple examples, by ensuring the example codes in documents produce an output inconsistent with the {code completion task} as mentioned in Section~\ref{subsubsec:constructionrag}.

\subsubsection{Retrieval Performance} As shown in Figure~\ref{fig:rq5retriever}, BM25 still outperforms GTE and performs on par with Text3. However, its Recall@5 rate drops a lot on such documents compared to the documents including the {target code completion task} as the example. 

\subsubsection{Generation Performance} Considering the low retrieval rate, we study the generation performance on setup \textit{w/\,tgt doc}. As shown in Table~\ref{tab:rq5-top1ondifftask}, RAG still effectively increases the pass rates of four LLMs from 0.4060$\sim$0.5564 (\textit{w/o doc}) to 0.6617$\sim$0.8346 (\textit{w/\,tgt doc}) on documents with examples mismatching the {code completion tasks}. 

However, we observe deleting examples (\textit{DelExmpl}) conversely enhances performance, notably for the less capable DSCoder and Qwen7B. Our analysis of failure cases reveals that LLMs often fail to identify differences between the {code completion tasks} and example codes and copy wrong usage patterns from the provided examples.

To understand whether example codes mismatching the {completion task} always hinder RAG performance, 
we first evaluate RAG on documents with only example code (\textit{OnlyExmpl}, \textit{i.e.}, removing description and parameter list from documents). Besides, we study the performance of DeepSeek-R1-Distill-Qwen-32B (\textit{abbr.} R1-Qwen32B), which shows exceptional programming ability through reasoning.\footnote{We didn't study R1-QWen32B in earlier RQs due to its much longer reasoning time and unique challenges, \textit{e.g.}, overthinking without producing final answers. Studying reasoning LLMs' behavior in our task remains an interesting future work.}
Results show that all five LLMs achieve higher pass rates on \textit{OnlyExmpl} than without document (\textit{w/o doc}). Besides, R1-QWen32B performs better on \textit{Original Doc} than on \textit{DelExmpl}. By studying the reasoning process of R1-Qwen32B, we notice that it demonstrates the ability to adapt example codes to target tasks. 

In fact, the mismatch between examples in documents and user goals has been identified as a hindrance for humans to learn API usage \cite{whyAPIhardtolearn,apilearning}. Our results reveal that \textit{\textbf{the issue also affects LLMs}}.

\subsubsection{Implications} 

Results confirm the usefulness of RAG on documents with examples mismatching the target task. Meanwhile, to enhance its effectiveness, we highly suggest RAG users illustrate more usage of each API by \textbf{introducing diverse example codes}, \textit{e.g.}, to cover more parameter combinations. Then, retrievers may be able to identify target APIs based on similar code contexts shared by example codes and {completion tasks}. 
Meanwhile, diverse examples may force the generator LLMs to distinguish usages in examples as they help humans \cite{whyAPIhardtolearn}. Besides, it is found that human developers can write a correct API usage by copying and adapting example codes \cite{apilearning}. Thus, enhancing reasoning ability may also help LLMs distinguish examples and learn API usage wisely.

{We demonstrate a task where new diverse examples help LLMs solve it. The task expects \code{y = ivy.conv1d\_transpose(x,filters,2,'SAME')}. Given the API description, parameter list, and a single example using \code{ivy.conv1d\_transpose(x,filters,1,'VALID',out=x)}, Qwen32B incorrectly generates \code{ivy.conv1d\_transpose(x,filters,strides=2,padding='VAL ID')} by copying ``\code{VALID}''. By adding four new examples where two use ``\code{VALID}'' and two use ``\code{SAME}'', Qwen32B successfully distinguishes between the two values and gives a correct solution \code{ivy.conv1d\_trans pose(x,filters,2,'SAME')}. This demonstrates the potential of diverse examples to help LLMs learn API usage wisely rather than directly copying.
Furthermore, initially, Qwen32B also fails to recognize \code{strides} as a positional-only parameter (already specified in the API description) and incorrectly sets it by name. This error is also fixed after new examples are given. This further demonstrates the help of diverse examples in guiding LLMs to learn API usage, which may also address the failures mentioned in RQ1 (Section~\ref{subsubsec:failurecases}).}
The complete task and document can be found in our artifact \cite{artifact}.

\section{Discussion}\label{sec:discussion}

\subsection{RAG for Code Generation from Natural Language Requirements}

\begin{table}[t]
\small
\caption{Pass Rates of RAG for Code Generation\label{tab:discussion-nlresult}}
\resizebox{0.99\linewidth}{!}
{
\setlength{\tabcolsep}{1.8pt}
\begin{tabular}{cccccc}
\toprule
\textbf{} & \textbf{\textit{w/o doc}} & {\begin{tabular}[c]{@{}c@{}}\textbf{\textit{w/\,top5 doc}}\\ \textit{(Original Doc)}\end{tabular}} & {\begin{tabular}[c]{@{}c@{}}\textbf{\textit{w/\,top5 doc}}\\ \textit{(DelExmpl)}\end{tabular}} & {\begin{tabular}[c]{@{}c@{}}\textbf{\textit{w/\,top5 doc}}\\ \textit{(DelParam)}\end{tabular}} & {\begin{tabular}[c]{@{}c@{}}\textbf{\textit{w/\,top5 doc}}\\ \textit{(DelDesc)}\end{tabular}} \\ \midrule
Qwen32B                                                 & 0.2697           & 0.4667                                                                       & 0.3394                                                                   & 0.4394                                                                   & 0.3879                                                                  \\
GPT-4o-mini                                                 & 0.2576           & 0.4667                                                                       & 0.3758                                                                   & 0.4364                                                                   & 0.3939                                                                  \\
Qwen7B                                                  & 0.1667           & 0.3939                                                                       & 0.2242                                                                   & 0.3364                                                                   & 0.2576                                                                  \\
DSCoder                                                & 0.1636           & 0.3818                                                                       & 0.1667                                                                   & 0.3636                                                                   & 0.2606                                                                  \\ 
\textit{(average)}                                                   & \textit{0.2144}           & \textit{0.4273}                                                                       & \textit{0.2765}                                                                   & \textit{0.3939}                                                                   & \textit{0.3250}                                                                  \\ 
\midrule
\begin{tabular}[c]{@{}c@{}}BM25\\ Recall@5\end{tabular} & -                & 0.3091                                                                       & 0.3273                                                                   & 0.2303                                                                   & 0.1515                                                                 \\ \bottomrule
\end{tabular}
}
\end{table}

In our study, we recommend API usage in the code completion setup. 
In this discussion, we further explore the usefulness of RAG in another LLM-assisted development scenario, code generation, where users provide a natural language (NL) requirement to prompt LLMs to generate whole code snippets. The exploration aims to broaden insights on RAG for unfamiliar API use. 

We did not find off-the-shelf NL requirements for programming tasks of unfamiliar APIs. Thus, we use GPT-4o to summarize an NL requirement for the example code (preprocessed as Section~\ref{subsubsec:defpbe} mentions) in API documents considering its effectiveness in code summarization \cite{prompt4oo1haodan}. The whole API document is given as context during summarization. The prompt is available in our artifact \cite{artifact}. 
We conduct the exploration on Polars and use BM25 retriever since it is also found effective in NL-based code generation \cite{empiricalofrag}.

As shown in Table~\ref{tab:discussion-nlresult}, using RAG increases the pass rates of four LLMs from 0.2144 (\textit{w/o doc}) to 0.4273 (\textit{w/\,top5 doc (Original Doc)}), demonstrating that RAG also enables LLMs to use unfamiliar APIs for code generation. Meanwhile, Recall@5 is only 0.3091, showing the known gap between NL requirement and API documentation \cite{empiricalofrag}. 
Deleting examples (\textit{Del Exmpl}) increases the Recall@5 from 0.3091 to 0.3273, but it decreases the final success rate from 0.4273 to 0.2765. This suggests that example code is vital for LLMs to learn API usage but may not benefit retrieval. \textit{DelParam} and \textit{DelDesc} hinder successful retrieval and decrease the final success rates.

To summarize, RAG also enables LLMs to use unfamiliar APIs for code generation. The generator LLMs still benefit from example codes; meanwhile, the retrievers face challenges such as bridging the semantic gap between NL requirements and API documents.

\subsection{Threats to Validity}

We identified potential threats to the validity of our study and have taken measures to mitigate them as follows:

\textbf{Representativeness of subject libraries.} This study explores the usefulness of RAG in recommending usages of APIs unfamiliar to LLMs. To balance the number of reliable documents and the unfamiliarity with LLMs, we collect four less common Python libraries used fairly infrequently in GitHub codes with four criteria. They are utilities of two major application domains of Python, data science and machine learning. They include 1017 eligible APIs.
We start with Python libraries following existing studies \cite{ccwanragstudy, awsragstudy, coderagbench}. 
Study results on these libraries show consistent conclusions with diversity (\textit{e.g.}, better robustness on Ivy). Besides, they reveal both consistent and unique findings compared to the popular Pandas library. Based on these, we consider our study results representative and able to generalize to the practical libraries unfamiliar to LLMs. 

\textbf{Representativeness of studied RAG setups.} We build a typical RAG pipeline with the popular \textit{langchain} \cite{langchain} framework \cite{SurveyonLLM}. RAG typically relies on a retriever and a generator LLM. We consider three popular retrievers found effective in existing studies \cite{coderagbench, awsragstudy, ccwanragstudy} and leaderboards \cite{mtebbench}. We study four LLMs from three families based on their outstanding performance on the BigCodeBench leaderboard \cite{BigCodeBench}. The setup should be representative.

\textbf{Representativeness of mutation.} We design mutation operators to reveal RAG performance on documents with varying quality. All operators are inspired by noise in API documents identified by existing studies as Section~\ref{subsec:mutoprs} introduces. We adopt GPT-4o to suggest synonyms and prepare new parameters. Despite the uncertainty of LLMs, GPT-4o is widely used in task preparation due to its strong natural language processing ability \cite{caofmbench, domaineval, selfinstruct, enhancingCLM, howfar}. Our manual check also suggests satisfying mutation results of GPT-4o. 

\section{Related Work}\label{sec:relatedwork}

\subsection{RAG for Code Generation and Completion}

Several studies focus on enhancing retrieval accuracy for better RAG performance.
\citet{KNN-TRANX} designed a tailored retriever aware of syntax constraints. \citet{LAIL} trained a retriever to select examples that the generator LLM prefers.
RepoCoder \cite{Repocoder} iteratively optimizes its retrieval results. 
ProCC \cite{ProCC} leverages three styles of code contexts to benefit retrieval.
In our study, enhancing retrieval accuracy also proves beneficial for RAG on API documentation as we discuss in RQ1. Enhancing retrieval of proper documents remains an interesting future work.

Another line of studies diversifies information sources for RAG, e.g., in-domain code bases \cite{Cocomic}, cross-code-file context \cite{Cocomic}, web resources \cite{su2024evor}, and Android development feature \cite{DroidCoder}.
\citet{RAR} generated code for low-resource languages based on language documentation and code examples, which act as a good knowledge base for uncommon domains. 
Unlike these works, we focus on RAG on API documentation to use less common APIs. In particular, we evaluate RAG in different setups. Our study results share experience in enabling LLMs to use unfamiliar library APIs.

\subsection{Empirical Study on RAG Performance}

To learn experience on the optimal setup for these essential configurations of RAG, researchers evaluated RAG under diverse document types \cite{cuconasu2024power}, document positions \cite{liu2024lost,cuconasu2024power}, and choice of retrievers \cite{wang2024searching}. 
There are also studies on the retrieval granularity \cite{chen2024dense}, length of contexts \cite{xu2023retrieval}, and integration with training-based methods \cite{ovadia2023fine}.
Besides configuration, \citet{rgb} evaluated RAG against noise and reveal that RAG may easily be misled by incorrect information in retrieved documents. 
These studies are mainly conducted on general applications like question answering.
Inspired by them, we explore RAG for API usage recommendation and share experience on enabling LLMs to code with less common library APIs. 

Recently, \citet{ccwanragstudy} studied the optimal setup of retrievers and prompts for question answering and code generation with popular libraries like Pandas. In comparison, we study the impact of API documentation quality on both the retrieval and generation phases. Besides, we focus on the less common libraries. We reveal both consistent helpfulness and unique challenges of RAG on the less common libraries and the popular Pandas library.

Besides, \citet{coderagbench} built CodeRAGBench to benchmark RAG on eight coding tasks. They mainly construct RAG databases with canonical solutions of coding tasks to explore the upper limits of RAG, as well as explore the limitation of RAG on open-domain resources. We follow their idea to benchmark RAG on the API documents that include the code completion task and its ground truth as examples in RQs1-4. With the setup, we identify the bottleneck in retrieval and LLM understanding ability. We also explore RAG performance on documentation with only an example mismatching the coding task in RQ5, and reveal the importance of diverse examples and reasoning ability to effectively guide LLMs.
\section{Conclusion}\label{sec:conclusion}

In this paper, we study an unexplored yet practical question: \textit{To what extent can RAG contribute when LLMs generate code using less common libraries?} To answer this question, we select four less common open-source Python libraries with a total of 1017 eligible APIs and conduct experiments on them. Our study yields several interesting findings, including the determining factors in improving LLMs' performance using RAG, the essential role of code examples to augment LLMs, and LLMs' noise-tolerant ability. Based on the results, we suggest developers pay more attention to the quality and diversity of the code examples in API documentation and enhance LLMs in reasoning to help them better learn API usage.

\balance

\bibliographystyle{ACM-Reference-Format}
\bibliography{references}

\end{CJK*}
\end{document}